\documentclass[final,onefignum,onetabnum]{siamonline181217}

\usepackage{lipsum}
\usepackage{amsfonts}
\usepackage{mathrsfs}
\usepackage{graphicx}
\usepackage{epstopdf}
\usepackage{enumerate}
\usepackage{subcaption}
\usepackage{algorithm}
\usepackage{algpseudocode}
\ifpdf
  \DeclareGraphicsExtensions{.eps,.pdf,.png,.jpg}
\else
  \DeclareGraphicsExtensions{.eps}
\fi

\newsiamremark{remark}{Remark}
\newsiamremark{hypothesis}{Hypothesis}
\crefname{hypothesis}{Hypothesis}{Hypotheses}
\newsiamthm{claim}{Claim}

\headers{Gaussian Process Landmarking for 3D GM}{T. Gao, S.Z. Kovalsky, D.M. Boyer, and I. Daubechies}

\title{Gaussian Process Landmarking for Three-Dimensional Geometric Morphometrics\thanks{Submitted
\funding{This work is supported by Simons Math+X Investigators Award 400837 and NSF CAREER Award BCS-1552848.}}}

\author{Tingran Gao\thanks{Computational and Applied
    Mathematics Initiative, Department of Statistics, The University of Chicago, Chicago IL (\email{tingrangao@galton.uchicago.edu})}
  \and Shahar Z. Kovalsky \thanks{Department of Mathematics, Duke University, Durham NC (\email{shaharko@math.duke.edu})}
  \and Doug M. Boyer \thanks{Department of Evolutionary Anthropology, Duke University, Durham NC (\email{doug.boyer@duke.edu}).}
\and Ingrid Daubechies \thanks{Department of Mathematics and Department of Electrical and Computer Engineering, Duke University, Durham NC (\email{ingrid@math.duke.edu})}}

\usepackage{amsopn}

\DeclareMathOperator*{\argmin}{argmin}
\DeclareMathOperator*{\argmax}{argmax}

\newcommand{\dd}{\,\mathrm{d}}

\newcommand{\GP}{\mathrm{GP}}

\usepackage[colorinlistoftodos]{todonotes}

\ifpdf
\hypersetup{
  pdftitle={Gaussian Process Landmarking for Three-Dimensional Geometric Morphometrics},
  pdfauthor={Tingran Gao, Shahar Z. Kovalsky, Doug M. Boyer, and Ingrid Daubechies}
}
\fi

\begin{document}

\maketitle

\begin{abstract}
  We demonstrate applications of the Gaussian process-based landmarking algorithm proposed in [T. Gao, S.Z. Kovalsky, and I. Daubechies, \emph{SIAM Journal on Mathematics of Data Science} (2019)] to geometric morphometrics, a branch of evolutionary biology centered at the analysis and comparisons of anatomical shapes, and compares the automatically sampled landmarks with the ``ground truth'' landmarks manually placed by evolutionary anthropologists; the results suggest that Gaussian process landmarks perform equally well or better, in terms of both spatial coverage and downstream statistical analysis. We provide a detailed exposition of numerical procedures and feature filtering algorithms for computing high-quality and semantically meaningful diffeomorphisms between disk-type anatomical surfaces.
\end{abstract}



\section{Introduction}

Computing one-to-one correspondences between objects, or \emph{registration}, plays an important role in a wide range of scientific disciplines, such as functional Magnetic Resonance Imaging (fMRI) in medical studies \cite{Pandey2009}, shape matching in computer graphics \cite{vanKaick2011}, remote sensing in geophysical sciences \cite{LeMoigne2017}, to name just a few. When the number of objects in comparison is prohibitively large, or when the information encoded in each data object is noisy or redundant, a common strategy is to reduce the comparison to \emph{alignment} or \emph{assignment} between sets of extracted \emph{features}, which are typically compact and informative representations of the original data constructed with domain knowledge. Examples of this sort include the \emph{scale-invariant feature transform} (SIFT) in computer vision \cite{Lowe1999} and \emph{anatomical landmarks} in statistical shape analysis \cite{DrydenMardia1998SSA} and computational anatomy \cite{JoshiMiller2000}. 

This paper stems from an attempt to apply principles of the statistics field of \emph{optimal experimental design} to automatically detect salient feature points on anatomical surfaces. For the biological morphometrical data of interest to us, these feature points surprisingly resemble the ``ground truth'' landmarks seasoned evolutionary anthropologists would identify as discrete anatomical points of correspondence. The proposed algorithm, based on successively selecting points of maximum \emph{mean squared prediction error} (MSPE) with respect to a Gaussian process modeled on a smooth manifold or from a scattered point cloud, turns out to be intimately connected with a class of greedy algorithms in \emph{reduced basis methods} \cite{BCDDvPW2011,DvPW2013} and enjoys fast rate of convergence. Interested readers may consult our companion paper \cite{GPLMK1} for more details on these theoretical aspects. This paper demonstrates that, not only do these Gaussian process landmarks perform equally well or better compared with the ground truth landmarks (in terms of both spatial coverage and downstream statistical analysis), they can also be leveraged to establish high-quality and semantically meaningful smooth bijections (registration maps) between biological structures; shape distances induced by these maps reach comparable explanatory power to ground truth landmarks for ordination (i.e., discriminating species groups) purposes.

The effectiveness of the proposed Gaussian process landmarking algorithm can be intuitively interpreted by the analogy between Gaussian process experimental design and the landmark selection procedure in geometric morphometrics. In a nutshell, the method we propose considers a Gaussian process on the anatomical surface, with a variance-covariance structure specified by the heat kernel of the underlying Riemannian manifold; the landmarks are then selected successively, each time picking a new landmark as the point on the surface with largest variance conditioned on all the previously selected landmarks. This procedure is highly reminiscent of the practice of \emph{landmarking} in \emph{geometric morphometrics}, the subfield of evolutionary biology focusing on quantifying the (dis-)similarities between pairs of two-dimensional anatomical surfaces based on their spatial configurations \cite{ZSS2012}.  The main objects of study in geometric morphometrics are anatomical surfaces, such as bones and teeth, of extinct and extant animals of particular interest to evolutionary biologists. The landmarking procedure typically starts with manually identifying an equal number of geometrically or semantically meaningful feature points, or \emph{landmarks}, on each specimen in a potentially large collection of anatomical surfaces; the landmarks are certified by domain experts to be in consistent one-to-one correspondences. The methodology of landmark selection, though difficult to articulate and is still constantly under debate (see e.g., very recent discussions \cite{WGGPS2018,Watanabe2018} and the references therein), emphasizes a comprehensive and balanced decision between sharp geometric features (points of application of real biomechanical forces) and points extremal to spatial configuration (points taken ``farthest away'' from other points under certain metrics). While geometric features could be recognized in a computational geometry manner by computing discretized curvatures, the precise meaning of ``extremality'' can be flexibly characterized as an ``uncertainly'' naturally encoded into the covariance structure of a Gaussian process with a geometry-aware kernel function. By carefully designing a curvature-reweighted heat kernel on discrete anatomical surfaces, the proposed algorithm is thus capable of selecting both sharp geometric features and extremal points in one single pass. Downstream Procrustes analysis \cite{Gower1975,DrydenMardia1998SSA,GowerDijksterhuis2004GPA} results presented in this paper also speak of the biological relevance of the automatically generated landmarks.



The rest of this paper is organized as follows. \Cref{sec:background} provides background materials on geometric morphometrics, Gaussian processes, and the Gaussian process landmarking algorithm proposed in \cite{GPLMK1}; \Cref{sec:theo-analysis} explains the importance of using a reweighted kernel in the Gaussian process landmarking algorithm for our application; \Cref{sec:application-autogm} details all the numerical procedures in this application; \Cref{sec:conclusions} closes with comments and discussions.

\section{Background}
\label{sec:background}

\subsection{Geometric Morphometrics: Old and New}
\label{sec:geom-morph}

Statistical shape analysis, often termed \emph{geometric morphometrics} in the context of comparative biology, is the quantitative analysis of variations and correlations among biological forms through the Cartesian coordinates of ``landmarks''---biologically informative, repeatable, and in some sense corresponding anatomical loci---on surfaces representing anatomy of biological organisms \cite{White2009,ARS2013,ZSS2012}. In order for the comparisons across specimens to be meaningful, practitioners in this field often require that the landmarks be consistently annotated on each specimen in a manner reflecting the ``operational homology\footnote{The term ``homology'' in the context of evolutionary theory bears a different meaning than in modern topology; see e.g., \cite[Part IV]{GMBlueBook1990}.}'' or ``biological correspondence'' (as discussed in \cite{Auto3dGM2015}) across individuated traits inherited from a common ancestry \cite{MG2009}. For instance, in the \emph{generalized Procrustes analysis} (GPA) framework \cite{Gower1975,DrydenMardia1998SSA,GowerDijksterhuis2004GPA,CP13}, the \emph{Procrustes distance} between two surfaces $S_1$, $S_2$ is computed using the following procedure:
\begin{enumerate}[(i)]
\item\label{item:3} Specify two sets of operationally homologous landmarks $\left\{ x_{\ell}^{\left( 1 \right)}\mid 1\leq \ell\leq L \right\}$ and $\left\{ x_{\ell}^{\left( 2 \right)}\mid 1\leq \ell\leq L \right\}$ on $S_1$ and $S_2$, respectively;
\item\label{item:4} Compute the distance between $S_1$ and $S_2$ by minimizing the energy functional
  \begin{equation}
    \label{eq:cP-variational}
    d_{\mathrm{cP}}\left( S_1,S_2 \right)=\inf_{T\in\mathbb{E}\left( 3 \right)}\left(\frac{1}{L}\sum_{\ell=1}^L \left\| T\left(x_{\ell}^{\left( 1 \right)}\right)-x_{\ell}^{\left( 2 \right)} \right\|^2\right)^{\frac{1}{2}}
  \end{equation}
where $\mathbb{E} \left( 3 \right)$ is the group of rigid motions in $\mathbb{R}^3$.
\end{enumerate}
This idea can be generalized to analyze a collection of consistently landmarked shapes, either assuming each set of landmarks on the same shape is centered at the origin so the variational problem is defined on a product space of orthogonal groups \cite{GowerDijksterhuis2004GPA,Nemirovski2007,So2011,NaorRegevVidick2013}, or estimate the optimal orthogonal and translation group elements jointly without the overall centering assumption \cite{CKS2015}.

The key to successfully applying the Procrustes framework in statistical shape analysis is to obtain an equal number of consistent, operationally homologous landmarks on every shape in a potentially enormous collection of shapes. Consistently landmarking a collection of shapes relies crucially upon domain knowledge and tedious manual labor, and the skill to perform it ``correctly'' typically requires years of professional training; even then the ``correctness'' can be subject to debate among experts (see e.g., \cite{PNAS2011} for an example on \emph{Lepilemur} teeth). In the first place, extracting a finite number of landmarks from a continuous surface inevitably loses geometric information, unless when the shapes under consideration are easily seen to be uniquely determined by the landmarks (e.g., planar polygonal shapes, as considered in \cite{Kendall1984}\cite{FryThesis1993}), which is rarely the case for geometric morphometricians in biology; this problem of ``inadequate coverage'' motivated the introduction of \emph{semilandmarks}---additional points along curves containing critical curvature information about the morphology---to compensate for the loss of geometry in the landmarking process. Unfortunately, the essential arbitrariness of the semilandmarks along a curve induces additional uncertainty that needs to be quantified and reduced \cite[\S 2]{ZSS2012}, especially in the absence of sharp anatomical features; the constraint of picking an equal number of landmarks on each shape also turns out to be far too artificial when the anatomical forms undergo complex evolutionary and developmental processes.

To mitigate both the scalability and the subjectivity issues in the existing Procrustes analysis framework, a recent trend of research in geometric morphometrics advocates \emph{automated} workflows to bypass the repetitive, laborious, and time-consuming process of manual landmark placement; see e.g., \cite{CP13,PNAS2011,PuenteThesis2013,LipmanDaubechies2011,Boyer2012,LipmanPuenteDaubechies2013,Auto3dGM2015,Gao2015Thesis,VMGBSB2017,KoehlHass2015,HLB2016} and the references therein. These techniques work directly with digitized anatomical surfaces represented as discrete triangular meshes; numerical algorithms are combined with computer graphics and geometry processing to provide high-throughput, landmark-free approaches for precise phenotyping \cite{PEKGJ2008,HKVHMJ2012,HSHRCKZEMS2014} on the discrete triangular meshes on their entirety (often consisting of thousands to millions of vertices in $\mathbb{R}^3$), without the manual landmarking stage to filter down the number of variables using potentially biased \emph{a priori} domain knowledge. Similar to the claim made by the proponents of landmark-based morphometrics that landmark coordinates contain more information than utilized in more traditional, measurement-based morphometrics \cite[\S 2]{ZSS2012}, the precursors of automated geometric morphometrics believe that using whole surfaces as input passes even more information to the downstream analysis.

Despite the capability of generating high quality pairwise shape registrations, automated geometric morphometric methods suffer from interpretability problems: since all comparisons are performed merely pairwise, composing the obtained correspondences along a closed loop does not give rise to an identity map in general. This lack of \emph{transitivity} (see e.g., \cite{GYDMB2017,HDM2016}) demands additional post-processing steps to translate the pairwise results into valid input for standard downstream phylogenetic analysis \cite{Paradis2011,PCM2014}, which bears a strong similarity with recent studies in \emph{synchronization problems} \cite{CKS2015,GBM2016,OSS2018}. The loop inconsistency of pairwise correspondences also challenges the interpretability of automated geometric morphometrics, since it becomes virtually impossible to identify functionally equivalent regions across distinct anatomical structures in a consistent manner. Until fully automated geometric morphometric algorithms reach the maturity with comparable explanatory power to a human practitioner of landmark-based geometric morphometrics, deeper and more systematic understanding of the landmarking process still seem of great interest and value.

The methodology we propose in this paper incorporates an algorithmic landmarking procedure into the automated pairwise registration algorithms. We point out that detecting morphometrically meaningful landmarks on anatomical surfaces in a completely unsupervised manner is generally a daunting task, since some of the most reliable landmarks are determined by patterns of juxtapositions of tissues---termed ``Type 1 landmarks'' by Bookstein \cite{Roth1993}---which are almost always absent on the triangular meshes input to automated algorithms \cite{Bookstein1991}. The selection process is further complicated by the requirement of the consistency of relative landmark positions across the data collected, as well as the specific functionality of the biological organism being studied \cite{ZSS2012}. While geometry processing algorithms (e.g., \cite{Banchoff1970,LLKR2007,CCFM2008,Bronstein2011,TCLLLMMSR2013}) are capable of detecting sharp geometric features characterized by metric or topology (Bookstein's ``Type 2 landmarks'' \cite{Roth1993}), as well as producing high quality pairwise registrations for accurate determination of operationally homologous loci, semilandmarks and a majority of ``Type 3 landmarks'' in Bookstein's typology of landmarks are marked simply for adequate and/or comprehensive coverage of the anatomical forms \cite{Roth1993}. For instance, some Type 3 landmarks are included in the analysis for being ``furthest away'' from sharp geometric or functional features \cite{ZSS2012}. These observations motivated us to consider algorithmic analogies of geometric morphologists' daily practice beyond the scope of computational geometry, shedding light upon landmark identification from the perspective of Bayesian statistics quantifying the ``uncertainty'' of morphometric analysis. As will be detailed in \Cref{sec:application-autogm}, we will first generate a set of candidate landmarks on each of the anatomical surfaces based on the ``uncertainty'' modeled by a Gaussian process,  then apply a matching scheme that filters out non-corresponding candidate landmarks between a pair of surfaces based upon bounded conformal distortion \cite{Lipman2012}.

\subsection{Gaussian Processes}
\label{sec:gaussian-processes}

A \emph{Gaussian process} (or \emph{Gaussian random field}) on a Polish space $M$ with mean function $m:M\rightarrow\mathbb{R}$ and covariance function $K:M\times M\rightarrow \mathbb{R}$ is defined as the stochastic process of which any finite marginal distribution on $n$ fixed points $x_1,\cdots,x_n\in M$ is a multivariate Gaussian distribution with mean vector
$$m_n:=\left( m \left( x_1 \right),\cdots,m \left( x_n \right) \right)\in\mathbb{R}^n$$
and variance-covariance matrix
\begin{equation}
  \label{eq:kernel-var-covar}
  K_n :=
\begin{pmatrix}
  K \left( x_1,x_1 \right) & \cdots & K \left( x_1,x_n \right)\\
  \vdots & & \vdots\\
  K \left( x_n,x_1 \right) & \cdots & K \left( x_n,x_n \right)
\end{pmatrix}\in\mathbb{R}^{n\times n}.
\end{equation}
A Gaussian process with mean function $m:M\rightarrow\mathbb{R}$ and covariance function $K:M\times M\rightarrow\mathbb{R}$ will be denoted as $\GP \left( m,K \right)$. Under model $Y\sim \GP \left( m,K \right)$, given observed values $y_1,\cdots,y_n$ at locations $x_1,\cdots,x_n$, the \emph{best linear predictor} (BLP) \cite{Stein2012,SWN2013} for the random field at a new point $x$ is given by the conditional expectation
\begin{equation}
\label{eq:blp}
  Y^{*}\left( x \right):=\mathbb{E}\left[ Y \left( x \right)\mid Y \left( x_1 \right)=y_1,\cdots,Y \left( x_n \right)=y_n \right]=m \left( x \right)+k_n \left( x \right)^{\top}K_n^{-1}\left( Y_n-m_n \right)
\end{equation}
where $Y_n=\left( y_1,\cdots,y_n \right)^{\top}\in \mathbb{R}^n$, $k_n \left( x \right)=\left( K \left( x,x_1 \right),\cdots,K \left( x,x_n \right) \right)^{\top}\in \mathbb{R}^n$; at any $x\in M$, the expected squared error, or \emph{mean squared prediction error} (MSPE), is defined as
\begin{equation}
  \label{eq:prediction-error}
  \begin{aligned}
    \mathrm{MSPE}\left( x \right):&=\mathbb{E}\left[ \left( Y \left( x \right)-Y^{*}\left( x \right) \right)^2 \right]\\
    &=\mathbb{E}\left[ \left(Y \left( x \right)-\mathbb{E}\left[ Y \left( x \right)\mid Y \left( x_1 \right)=y_1,\cdots,Y \left( x_n \right)=y_n \right]\right)^2 \right]\\
    &=K \left( x,x \right)-k_n \left( x \right)^{\top}K_n^{-1}k_n \left( x \right)
  \end{aligned}
\end{equation}
which is a function over $M$. Here the expectation is with respect to all realizations $Y\sim\GP \left( m,K \right)$. Squared integral ($L^2$) or sup ($L^{\infty}$) norms of the pointwise MSPE are often used as a criterion for evaluating the prediction performance over the experimental domain. In geospatial statistics, interpolation with \eqref{eq:blp} and \eqref{eq:prediction-error} is known as \emph{kriging}.

\subsection{Gaussian Process Landmarking}
\label{sec:gauss-proc-landm}

We now provide a quick summary of the \emph{Gaussian process landmarking} algorithm from \cite{GPLMK1}. The input to this algorithm is a triangular mesh $G=\left( V,E,F \right)$. Denote the set of vertices $V$ as $\left\{ x_1,\cdots,x_{\left| V \right|}\right\}\subset\mathbb{R}^3$, and construct the \emph{discrete heat kernel}
\begin{equation}
\label{eq:sq-exp-kernel-submfld}
  K = \left( K_{ij} \right)_{1\leq i,j\leq n} := \left[ \exp \left( -\frac{\left\| x_i-x_j \right\|^2}{t} \right) \right]_{1\leq i,j\leq n}.
\end{equation}
We calculate the mean and Gaussian curvature functions $\eta:V\rightarrow \mathbb{R}$, $\kappa:V\rightarrow\mathbb{R}$ on the triangular mesh $\left( V,E,F \right)$ using standard algorithms in computational differential geometry \cite{CM2003,ACDLD2003}. On a two-dimensional surface $S$, Gaussian curvature $\kappa \left( x \right)$ at a point $x\in S$ is the product of the two principal curvatures $k_1 \left( x \right)$, $k_2 \left( x \right)$ at $x$, i.e., $\kappa \left( x \right)=k_1 \left( x \right)k_2 \left( x \right)$, and the mean curvature is defined as $\eta \left( x \right)=k_1 \left( x \right)+k_2 \left( x \right)$. For any $\lambda\in \left[ 0,1 \right]$ and $\rho>0$, define the value of the weight function $w=w_{\lambda,\rho}$ at each vertex $x_i$ by
\begin{equation}
  \label{eq:weight-function-discrete}
  w_{\lambda,\rho}\left( x_i \right)=\frac{\lambda \left| \kappa \left( x_i \right) \right|^{\rho}}{\displaystyle \sum_{k=1}^{\left| V \right|} \left| \kappa \left( x_k \right) \right|^{\rho}\nu \left( x_k \right)}+\frac{\left( 1-\lambda \right) \left| \eta \left( x_i \right) \right|^{\rho}}{\displaystyle \sum_{k=1}^{\left| V \right|} \left| \eta \left( x_k \right) \right|^{\rho}\nu \left( x_k \right)},\qquad \forall x_i\in V
\end{equation}
where $\nu \left( x_k \right)$ is the area of the Voronoi cell centered at $x_i$ on the triangular mesh $T$, $W$ is a diagonal matrix of size $n\times n$ with $w \left( x_k \right)\nu \left( x_k \right)$ at its $k$-th diagonal entry ($1\leq k\leq \left| V \right|$), and $K$ is the discrete squared exponential kernel matrix \cref{eq:sq-exp-kernel-submfld}. The parameter $\lambda$ controls the relative proportion of mean and Gaussian curvatures, and the parameter $\rho$ adjusts the sharpness of the energy surfaces of mean and Gaussian curvatures before they are combined to form the weight function. Denoting $W$ for the diagonal matrix with weight $w_{\lambda,\rho}\left( x_k \right)$ at the $k$th diagonal element, we define the \emph{reweighted heat kernel} $k_t^{w_{\lambda,\rho}}$ on $V\times V$ as
\begin{equation}
\label{eq:location-prior-kernel-discrete}
  k_t^{w_{\lambda,\rho}}\left( x_i,x_j \right)=\sum_{k=1}^{\left| V \right|}k_{t/2}\left( x_i,x_k \right)k_{t/2}\left( x_k,x_j \right)w_{\lambda,\rho}\left( x_k \right)\nu \left( x_k \right)=:\left(\tilde{K}^{\top}W\tilde{K}\right)_{ij}
\end{equation}
where the unweighted kernel $k_{t/2}$ is calculated as in \cref{eq:sq-exp-kernel-submfld} but with bandwidth parameter $t/2$ instead of $t$, i.e.,
\begin{equation}
\label{eq:K-tilde}
  \tilde{K}:=\left( e^{-\frac{\left\| x_i-x_j \right\|^2}{t/2}} \right)_{1\leq i,j\leq \left| V \right|}\in\mathbb{R}^{\left| V \right|\times \left| V \right|}.
\end{equation}
Note that, even for the trivial weight function $w\equiv 1$ (i.e., when $W$ is the identity matrix of size $\left| V \right|\times \left| V \right|$), \cref{eq:location-prior-kernel-discrete} does not reduce to \cref{eq:sq-exp-kernel-submfld} since it is easy to check that $\tilde{K}^2\neq K$; this is in stark contrast with the reproducing case in the continuous regime in \cite[(2.8)]{GPLMK1}. Note only are heat kernels natural choices for specifying Gaussian processes on Riemannian manifolds (see e.g., a brief survey in \cite[\textsection 2.1]{GPLMK1}), their infinite smoothness also ensures fast convergence of the proposed Gaussian process landmarking algorithm; see \cite[Theorem 4.6]{GPLMK1}.

We now detail the greedy landmarking procedures. Until a fixed total number of landmarks are collected, at step $\left( n+1 \right)$ the algorithm computes the uncertainty score $\Sigma_{\left( k \right)}$ on $V$ from the existing $n$ landmarks $\xi_1,\cdots,\xi_n$  by
\begin{equation}
\label{eq:uncertainty-score}
\begin{aligned}
  \Sigma_{\left( n+1 \right)} \left( x_i \right)=k_t^{w_{\lambda,\rho}}\left( x_i, x_i \right)-k_t^{w_{\lambda,\rho}} \left( x_i,\xi_n^1 \right)^{\top}
  k_t^{w_{\lambda,\rho}}\left( \xi_n^1,\xi_n^1 \right)^{-1}k_t^{w_{\lambda,\rho}} \left( x_i,\xi_n^1 \right)
\end{aligned}
\end{equation}
for all $x_i\in V$, where
\begin{equation*}
  \begin{aligned}
    k_t^{w_{\lambda,\rho}} \left( x_i,\xi_n^1 \right)&:=
   \left( k_t^{w_{\lambda,\rho}}\left( x_i,\xi_1 \right),\cdots,k_t^{w_{\lambda,\rho}}\left( x_i,\xi_n \right)\right)^{\top},\\
  k_t^{w_{\lambda,\rho}} \left( \xi_n^1,\xi_n^1 \right)&:=\begin{pmatrix}
    k_t^{w_{\lambda,\rho}}\left( \xi_1,\xi_1 \right) & \cdots & k_t^{w_{\lambda,\rho}}\left( \xi_1,\xi_n \right)\\
    \vdots & & \vdots\\
    k_t^{w_{\lambda,\rho}}\left( \xi_n,\xi_1 \right) & \cdots & k_t^{w_{\lambda,\rho}}\left( \xi_n,\xi_n \right)
  \end{pmatrix},
  \end{aligned}
\end{equation*}
and pick the $\left( n+1 \right)$-th landmark $\xi_{n+1}$ according to the rule
\begin{equation*}
  \xi_{n+1} = \argmax_{x_i\in V}\Sigma_{\left( n+1 \right)}\left( x_i \right).
\end{equation*}
If there exist more than one maximizers of $\Sigma_{\left( n+1 \right)}$, we just randomly pick one; at step $1$ the algorithm simply picks the vertex maximizing $x\mapsto k_t^{w_{\lambda,\rho}}\left( x,x \right)$ on $V$. See \cref{alg:gaussian-process-landmarking} for a comprehensive description. We stick in this paper with default values $\lambda=1/2$ and $\rho=1$ in all examples and applications, but one may wish to alter these values to fine-tune the landscape of the weight function for a specific application. The bandwidth parameter $t$ is set to be the average edge length of the triangular mesh, following standard practices of manifold learning (see e.g., \cite{CoifmanLafon2006}). More details can be found in \cite[\S3]{GPLMK1}; we only mention here that \cref{eq:weight-function-discrete} and \cref{eq:location-prior-kernel-discrete} are discretizations of
\begin{equation}
  \label{eq:curvature-weight-function}
  w_{\lambda,\rho} \left( x \right) = \frac{\lambda\left| \kappa \left( x \right) \right|^{\rho}}{\displaystyle \int_M \left|\kappa \left( \xi \right)\right|^{\rho}\dd\mathrm{vol}_M \left( \xi \right)}+\frac{\left( 1-\lambda \right)\left| \eta \left( x \right) \right|^{\rho}}{\displaystyle \int_M \left|\eta \left( \xi \right)\right|^{\rho}\dd\mathrm{vol}_M \left( \xi \right)},\quad\forall x\in M
\end{equation}
and
\begin{equation}
  \label{eq:location-prior-kernel}
  k^w_t \left( x,y \right)=\int_Mk_{t/2}\left( x,z \right)k_{t/2} \left( z,y \right)w \left( z \right)\dd \mathrm{vol}_M \left( z \right),\quad\forall x,y \in M
\end{equation}
respectively. Since \cref{alg:gaussian-process-landmarking} builds upon the heat kernel and curvature of the underlying Riemannian manifold, the landmarks are stable under isometric deformations; we expect them to be stable across near-isometric Riemannian manifolds as well, which are objects of main interest to applications in geometric morphometrics.

\begin{figure}[htb]
  \centering
  \includegraphics[width=1.0\textwidth]{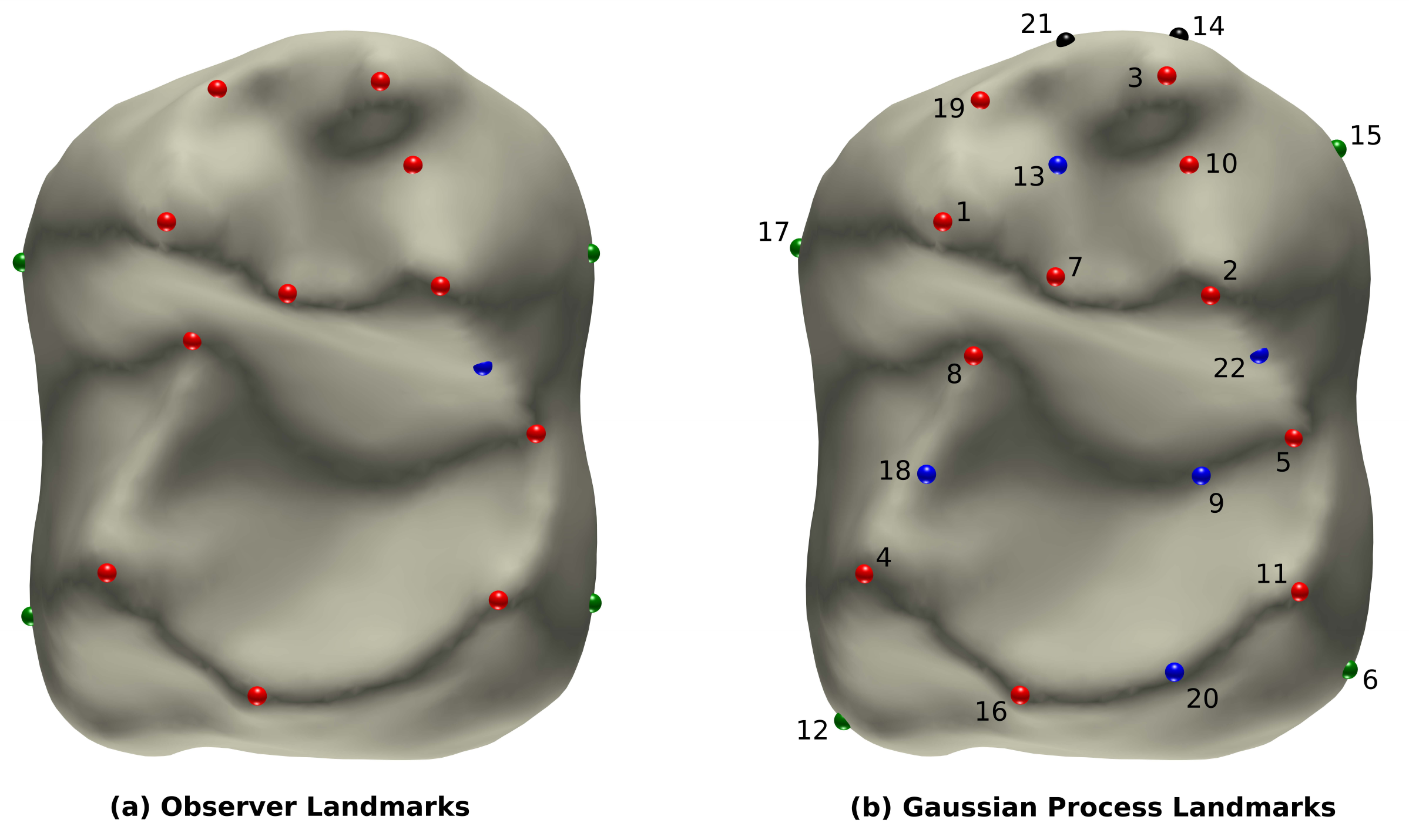}
  \caption{\small\textbf{Left:} Observer landmarks on a digitized fossil molar of a \emph{Teilhardina} (one of the oldest known fossil primates closely related with living tarsiers and anthropoids \cite{Teilhardina2017}) identified manually by an evolutionary anthropologist as ground truth, first published in \cite{PNAS2011}. \textbf{Right:} Landmarks picked by Gaussian Process Landmarking (\cref{alg:gaussian-process-landmarking}). The numbers next to each landmark indicate the order of appearance. These automatically computed landmarks strikingly resemble the observer landmarks; the 22 first Gaussian process landmarks already capture the same features selected by an expert anthropologist: the red landmarks (Number $1$-$5$, $7$, $8$, $10$, $11$, $16$, $19$) signal sharp geometric features (cusps or saddle points corresponding to local maximum/minimum Gaussian curvature); the blue landmarks sit either along the curvy cusp ridges and grooves (Number $13$, $18$, $20$, $22$) or at the basin (Number $9$), serving the role often played by semilandmarks; the four green landmarks (Number $6$, $12$, $15$, $17$) approximately delimit the ``outline'' of the tooth in occlusal view. The ordering of the Gaussian process landmarks does not affect the bounded distortion surface registration algorithm in \Cref{sec:surf-registr-match}.}
  \label{fig:Q10-ObLmk-vs-GPLmk}
\end{figure}

Guided by uncertainty specified through such curvature-reweighted covariance structure, the Gaussian process landmarking often identifies landmarks of abundant biological information---for instance, the first few Gaussian process landmarks are often highly biologically informative, and demonstrate comparable level of coverage with the observer landmarks manually picked by human experts. See \cref{fig:Q10-ObLmk-vs-GPLmk} for a visual comparison between the automatically generated landmarks with the observer landmarks manually placed by evolutionary anthropologists on a different digitized fossil molar. 

\begin{algorithm}
\caption{{\sc Gaussian Process Landmarking with Reweighted Heat Kernel}}
\label{alg:gaussian-process-landmarking}
 \begin{algorithmic}[1]
   \Procedure{GPL}{$T$, $L$, $\lambda\in \left[ 0,1 \right]$, $\rho>0$, $\epsilon>0$}\Comment{Triangular Mesh $T =\left( V,E \right)$, number of landmarks $L$}
   \State{$\kappa,\eta\gets$ {\sc DiscreteCurvatures}$\left( T \right)$}\Comment{calculate discrete Gaussian curvature $\kappa$ and mean curvature $\eta$ on $T$}
   \State{$\nu\gets$ {\sc VoronoiAreas}$\left( T \right)$}\Comment{calculate the area of Voronoi cells around each vertex $x_i$}
   \State{$w_{\lambda,\rho}\gets$ {\sc CalculateWeight}$\left(\kappa,\eta,\lambda,\rho,\nu\right)$}\Comment{calculate weight function $w_{\lambda,\rho}$ according to \cref{eq:curvature-weight-function}}
   \State{$W\gets\left[ \exp\left(-\left\| x_i-x_j \right\|^2\bigg/\epsilon\right) \right]_{1\leq i,j\leq \left| V \right|}\in\mathbb{R}^{\left| V \right|\times \left| V \right|}$}
   \State{$\Lambda\gets\mathrm{diag}\left( w_{\lambda,\rho}\left( x_1 \right)\nu \left( x_1 \right), \cdots, w_{\lambda,\rho}\left( x_{\left| V \right|} \right)\nu \left( x_{\left| V \right|} \right)\right)\in\mathbb{R}^{\left| V \right|\times \left| V \right|}$}
   \State{$\xi_1,\cdots,\xi_L\gets\emptyset$}\Comment{initialize landmark list}
   \State{$\Psi\gets 0$}
   \State{$\ell\gets 1$}
   \State{$K_{\mathrm{full}}\gets W^{\top}\Lambda W\in\mathbb{R}^{\left| V \right|\times \left| V \right|}$}
   \State{$K_{\mathrm{trace}}\gets \mathrm{diag}\left(K_{\mathrm{full}}\right)\in\mathbb{R}^{\left| V \right|}$}
   \While{$\ell < L+1$}
     \If {$\ell=1$}
       \State $\Sigma\gets K_{\mathrm{trace}}$
     \Else
       \State $\Sigma\gets K_{\mathrm{trace}}-\textrm{diag}\left(\Psi^{\top} \left(\Psi \left[\left[\xi_1,\cdots,\xi_{\ell}\right],: \right]\big\backslash\Psi\right)\right)\in\mathbb{R}^{\left| V \right|}$\Comment{calculate uncertainty scores by \cref{eq:uncertainty-score}}
     \EndIf
     \State $\displaystyle\xi_{\ell}\gets \argmax\Sigma$
     \State $\Psi\gets K_{\mathrm{full}} \left[ :,\left[\xi_1,\cdots,\xi_{\ell}\right] \right]$
     \State $\ell\gets \ell+1$
   \EndWhile
 \State \Return $\xi_1,\cdots,\xi_L$
   \EndProcedure
 \end{algorithmic}
\end{algorithm}

\Cref{alg:gaussian-process-landmarking} greedily picks as the next landmark the point on the manifold $M$ that maximizes the pointwise prediction error. This sequential optimization approach is reminiscent of a popular approximation scheme in entropy-based experimental design \cite{KSG2008,SKKS2010,SWN2013}. The key observation here is the equivalence between minimizing the \emph{conditional entropy} (or maximizing the \emph{information gain}) and maximizing the determinant of the marginal covariance matrix at the design points; see e.g., \cite[\S6.2.1]{SWN2013}, \cite[\S3.1]{KSG2008}, \cite[\S2.2]{SKKS2010}. More concretely, if we denote the maximum entropy of a Gaussian process $\GP \left( m,K \right)$ on a manifold $M$ with respect to any $n$ observations as
\begin{equation*}
  \mathrm{OPT}\left( n \right):=\max_{\left\{x_1,\cdots,x_n\right\}\subset M} \det K_n
\end{equation*}
where $K_n$ is the variance-covariance matrix defined in \cref{eq:kernel-var-covar}, and write $\mathrm{GPL}\left( n \right)$ for the value of $\det K \left( X_1,\cdots,X_n \right)$ where $X_1,\cdots,X_n$ are generated using \cref{alg:gaussian-process-landmarking}, then by the \emph{submodularity} of the entropy function \cite{KLQ1995} we conclude, using classic results \cite{NWF1978}, that
\begin{equation*}
  \mathrm{OPT}\left( n \right)\geq\mathrm{GPL}\left( n \right)\geq \left( 1-1/e \right)\mathrm{OPT}\left( n \right).
\end{equation*}
In other words, the entropy of the greedy algorithm is equivalent to the maximum entropy up to a multiplicative constant. While this general submodularity-based framework justifies the information-theoretic asymptotic near-optimality of Gaussian Process Landmarking (\cref{alg:gaussian-process-landmarking}), in \cite{GPLMK1} we established a stronger optimality result in the context of Gaussian processes: up to a multiplicative constant, the MSPE with respect to $n$ Gaussian process landmarks coincides with the optimal MSPE attainable over all sets of $n$ points on $M$, for any positive integer $n\in\mathbb{Z}_{\geq 0}$. More details about this can be found in \cite[\S4]{GPLMK1}.

\section{The Role of Reweighting in the Kernel Construction}
\label{sec:theo-analysis}

An essential construction in the application of \cref{alg:gaussian-process-landmarking} to geometric morphometrics is the reweighted kernel \cref{eq:location-prior-kernel}. Intuitively, the reweighting step modifies the Euclidean heat kernel \cref{eq:sq-exp-kernel-submfld} by amplifying the influence of locations with relatively high weights. To investigate the role played by the reweighted kernel in greater detail, we study the behavior of the reweighted kernel in the asymptotic regime when the number of samples increases to infinity. As will be demonstrated in \cref{thm:ptwise-convergence}, the reweighted kernel defines a diffusion process on the manifold with a backward Kolmogorov operator conjugate to the Witten Laplacian, of which the eigenfunctions has desired localization behavior near critical points of the weight function.

To motivate \cref{thm:ptwise-convergence}, let us consider i.i.d. (with respect to the standard volume measure) samples $\left\{ x_i \right\}_{i=1}^N$ on a closed Riemannian manifold $\left(M,g\right)$, as well as an arbitrary function $f\in C^2 \left( M \right)$. To simplify the discussion, assume for the moment that the samples are uniformly distributed on $M$ with respect to the normalized volume form on $M$. It is well known in manifold learning (see e.g., \cite{Singer2006ConvergenceRate,CoifmanLafon2006}) that, for any $x\in M$,
\begin{equation}
\label{eq:lln}
  \frac{1}{N}\sum_{j=1}^N e^{-\frac{\left\| x-x_j \right\|^2}{2\epsilon}} f \left( x_j \right)\longrightarrow \frac{1}{\mathrm{Vol}\left( M \right)}\int_M e^{-\frac{\left\| x-y \right\|^2}{2\epsilon}} f \left( y \right)\dd \mathrm{Vol}\left( y \right)
\end{equation}
as $n\rightarrow\infty$. Denote the positive weight function \cref{eq:curvature-weight-function} used in \cref{eq:location-prior-kernel} by
\begin{equation}
\label{eq:weight-function-exponential-form}
  w \left( x \right)=e^{-V \left( x \right)}, \quad V \left( x \right)\geq 0 \qquad \forall x\in M.
\end{equation}
Note that the normalization in \cref{eq:curvature-weight-function} ensures that the value of the weight function is bounded between $0$ and $1$, which is implied in \cref{eq:weight-function-exponential-form}, for numerical stability. Repeatedly using the ``law of large numbers'' argument \cref{eq:lln}, we have the following convergence for the reweighted kernel:
\begin{equation*}
  \begin{aligned}
    \frac{1}{N^2}&\sum_{k=1}^N e^{-\frac{\left\| x-x_k \right\|^2}{2\epsilon}} e^{-V \left( x_k \right)} \sum_{j=1}^N e^{-\frac{\left\| x_k-x_j \right\|^2}{2\epsilon}} f \left( x_j \right)\longrightarrow\\
    &\frac{1}{\left[ \mathrm{Vol}\left( M \right) \right]^2}\int_M\!\int_M e^{-\frac{\left\| x-z \right\|^2}{2\epsilon}} e^{-V \left( z \right)} e^{-\frac{\left\| z-y \right\|^2}{2\epsilon}} f \left( y \right)\dd\mathrm{Vol}\left( z \right)\dd\mathrm{Vol}\left( y \right)\quad\textrm{as $N\rightarrow\infty$.}
  \end{aligned}
\end{equation*}
This motivates us to consider the asymptotic behavior of the integral operator on the right hand side in the asymptotic regime $\epsilon\rightarrow\infty$, a recurring theme in manifold learning (see e.g., \cite{LapEigMaps2003,CoifmanLafon2006,SingerWu2012VDM,HDM2016}). For simplicity of notation, we will denote the integral operator with respect to kernel $K:M\times M\rightarrow\mathbb{R}$ as
\begin{equation*}
  T_Kf \left( x \right):=\frac{1}{\left| \mathrm{Vol}\left( M \right) \right|}\int_MK \left( x,y \right)f \left( y \right)\,\mathrm{dVol}\left( y \right).
\end{equation*}

\begin{theorem}
\label{thm:ptwise-convergence}
Let $M$ be a $d$-dimensional closed manifold. For any smooth functions $f,V\in C^{\infty} \left( M \right)$ with  $V\geq 0$, for kernel function
\begin{equation}
  \label{eq:reweighted-kernel}
  K_V^{\epsilon} \left( x,y \right):=\frac{1}{\left| \mathrm{Vol}\left( M \right) \right|}\int_M e^{-\frac{\left\| x-z \right\|^2}{2\epsilon}} e^{-V \left( z \right)} e^{-\frac{\left\| z-y \right\|^2}{2\epsilon}}\dd\mathrm{Vol}\left( z \right)
\end{equation}
we have
\begin{equation}
\label{eq:pointwise-concergence-formula}
  \frac{T_{K^{\epsilon}_V}f \left( x \right)}{T_{K^{\epsilon}_V}1}\rightarrow f \left( x \right)+\epsilon \Big(\Delta f \left( x \right) - \nabla f \left( x \right)\cdot \nabla V \left( x \right)\Big)+O \left( \epsilon^{\frac{3}{2}} \right)\quad\textrm{as$\,\,\epsilon\rightarrow 0$}
\end{equation}
where $1$ stands for the constant function on $M$ taking value $1$.
\end{theorem}

The proof of \cref{thm:ptwise-convergence} can be found in \Cref{sec:proof-theorem-ptws-convergence}. \Cref{thm:ptwise-convergence} indicates that a proper normalization of the reweighted kernel \cref{eq:location-prior-kernel} gives rise to an approximation to the heat kernel of the backward Kolmogorov operator
$$L=-\Delta+\nabla V\cdot \nabla.$$
Note that this operator is also the infinitesimal generator of the diffusion process determined by the stochastic differential equation
\begin{equation*}
  \dd X_t=-\nabla V \left( X_t \right)\dd t+\sqrt{2}\dd W_t
\end{equation*}
where $W_t$ is the standard Wiener process defined on $M$. In particular, \cref{thm:ptwise-convergence} suggests that our construction of the reweighted kernel encodes information from the weight function into the dynamics of the stochastic process on $M$.

\begin{remark}
It is interesting to compare the result of \cref{thm:ptwise-convergence} with the related but different kernel construction in diffusion maps \cite{CoifmanLafon2006,NLCK2006}. The integral operator in \cref{eq:pointwise-concergence-formula} is a properly normalized version of integrating a smooth function against the kernel
\begin{equation*}
  \widetilde{K}_V\left( x,y \right):=\int_M e^{-\frac{\left\| x-z \right\|^2}{2\epsilon}}e^{-V \left( z \right)} e^{-\frac{\left\| z-y \right\|^2}{2\epsilon}} \dd\mathrm{Vol}\left( z \right)
\end{equation*}
which is obtained from sandwiching $\exp \left[ -V \left( \cdot \right) \right]$ with the squared exponential kernel. If we pick a different order of sandwiching, namely, construct the kernel as
\begin{equation*}
  K_V \left( x,y \right):= e^{-V \left( x \right)} e^{-\frac{\left\| x-y \right\|^2}{2\epsilon}} e^{-V \left( y \right)},
\end{equation*}
then a quick computation in the same spirit as \cite[Theorem 1]{BerryHarlim2016} leads to
\begin{equation*}
    \frac{T_{K_V}f \left( x \right)}{T_{K_V}1}\rightarrow f \left( x \right)+\epsilon \left( \frac{1}{2}\Delta f \left( x \right)-\nabla f \left( x \right)\cdot\nabla V \left( x \right) \right)+O \left( \epsilon^{\frac{3}{2}} \right)\quad\textrm{as$\,\,\epsilon\rightarrow 0$.}
\end{equation*}
In other words, the infinitesimal generator differs from the one calculated in \cref{eq:pointwise-concergence-formula} only by a multiplicative factor in front of the Laplace-Beltrami operator. We actually observe very similar numerical results when replacing $\widetilde{K}_V$ with $K_V$ in \cref{alg:gaussian-process-landmarking}, when appropriate bandwidth parameters are chosen. Empirically, we find the bandwidth parameter easier to tune for kernel $\widetilde{K}_V$ than for $K_V$.  
\end{remark}

With these preparation, we now illustrate the effect of kernel reweighting by replacing \cref{eq:weight-function-exponential-form} with
\begin{equation}
\label{eq:weight-function-exponential-form-variable-bandwidth}
w_{\epsilon} \left( x \right)=e^{-V \left( x \right)/\epsilon}, \quad V \left( x \right)\geq 0 \qquad \forall x\in M.
\end{equation}
Note that we can always redefine the potential function $V$ so that \cref{eq:weight-function-exponential-form} is written in the form of \cref{eq:weight-function-exponential-form-variable-bandwidth}, for any fixed $\epsilon>0$; in particular, replacing $V \left( x \right)$ by $V \left( x \right)/\epsilon$ does not change its critical points. An almost identical calculation as in the proof of \cref{thm:ptwise-convergence} leads to

\begin{figure}[htp]
  \centering
  \includegraphics[width=1.0\textwidth]{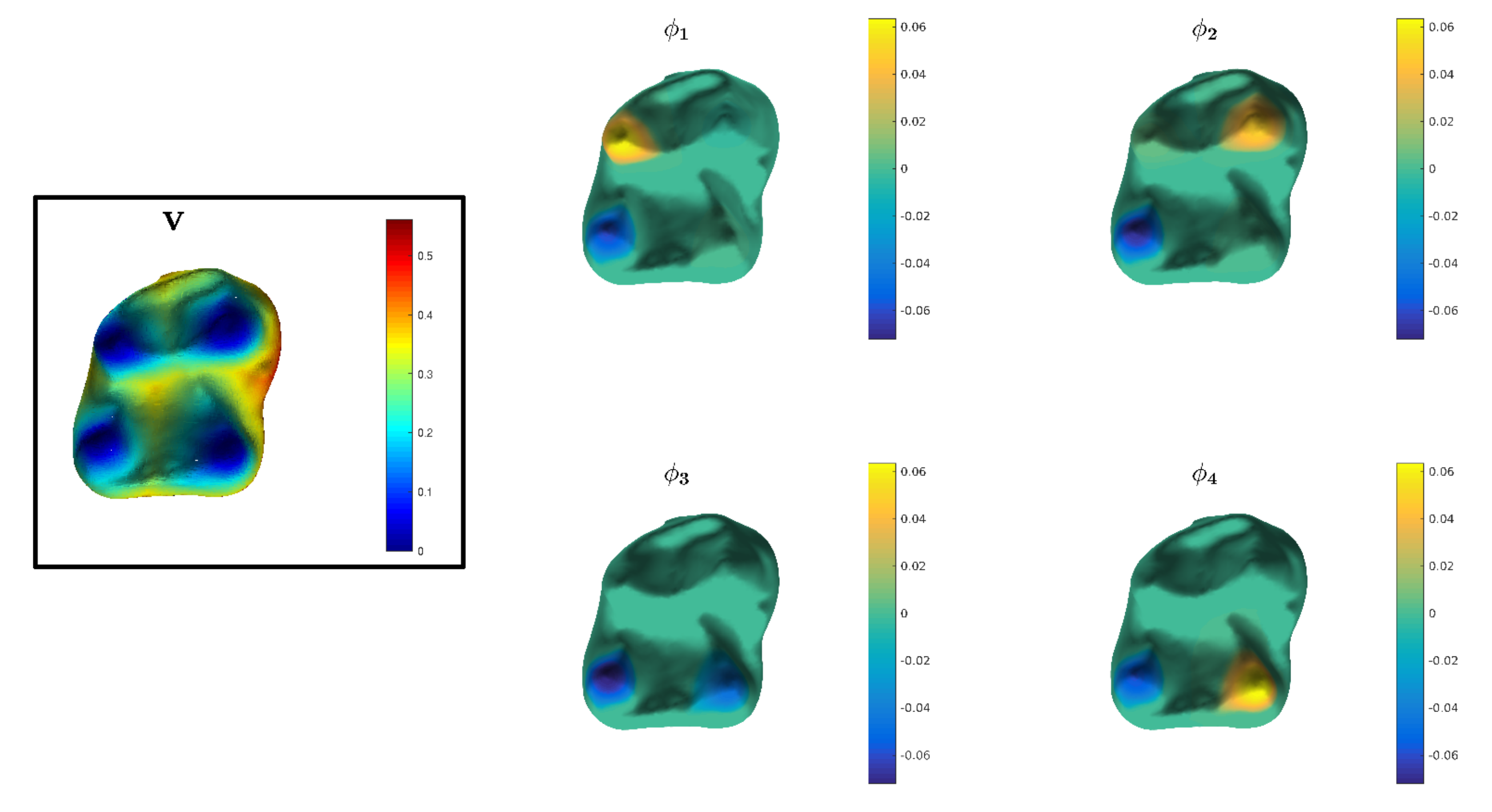}
  \caption{\small Concentration of the eigenfunctions of the Witten Laplacian with respect to a potential function $V$. The left boxed subplot illustrates the potential function $V$ using a heat map; we constructed this potential to have $4$ wells centered around $4$ cusps manually picked on a \emph{Lorisidae} mandibular molar. The $4$ eigenfunctions corresponding to the $4$ smallest eigenvalues of the Witten Laplacian with potential $V$ are depicted on the $2\times 2$ panel on the right; it can be read off from the colorbars that the support of each of these $4$ eigenfunctions are concentrated around critical points of $V$.}
  \label{fig:potential-j09}
\end{figure}

\begin{corollary}
\label{cor:varying-scaled-gradient-potential}
  Under the same assumptions as in \cref{thm:ptwise-convergence},
\begin{equation}
\label{eq:pointwise-concergence-formula-semi-classical}
  \frac{T_{K_{V/\epsilon}^{\epsilon}}f \left( x \right)}{T_{K_{V/\epsilon}^{\epsilon}}1}\rightarrow f \left( x \right)+\epsilon \Big(\Delta f \left( x \right) - \nabla f \left( x \right)\cdot \frac{1}{\epsilon}\nabla V \left( x \right)\Big)+O \left( \epsilon^{\frac{3}{2}} \right)\quad\textrm{as$\,\,\epsilon\rightarrow 0$}
\end{equation}
where $K^{\epsilon}_V/\epsilon$ is obtained from \cref{eq:reweighted-kernel} by replacing the potential $V$ with $\frac{1}{\epsilon}V$.
\end{corollary}
Similar to \cref{thm:ptwise-convergence}, \cref{cor:varying-scaled-gradient-potential} shows that a proper normalization of the kernel reweighted by \cref{eq:weight-function-exponential-form-variable-bandwidth} approximates the heat kernel of the operator
$$L_{\epsilon}:=-\Delta+\frac{1}{\epsilon}\nabla V\cdot \nabla.$$
The dependence on $\epsilon$ is of particular interest. It is well known in the literature of semi-classical analysis that $\epsilon L_{\epsilon}$ is conjugate to the \emph{semiclassical Witten Laplacian}~\cite{Witten1982} on $0$-forms:
\begin{equation*}
  \Delta_{V,\epsilon}=\epsilon e^{-V/2\epsilon}\left( \epsilon L_{\epsilon} \right)e^{V/2\epsilon}
\end{equation*}
where
\begin{equation*}
  \Delta_{V,\epsilon}=-\epsilon^2\Delta+\frac{1}{4}\left| \nabla V \right|^2-\frac{\epsilon}{2}\Delta V.
\end{equation*}
Since the eigenfunctions of $\Delta_{V,\epsilon}$ corresponding to the leading small eigenvalues concentrate near the critical points of the potential function $V$ for sufficiently small $\epsilon>0$ (see e.g., \cite{HS1985} or \cite[Theorem 3.9]{LPNV2013} for mathematically more precise statements), the eigenfunctions corresponding to the leading small eigenvalues of $L_{\epsilon}$ also concentrate near the critical points of $V$ after being multiplied by $e^{-V/2\epsilon}$, which, by the proof of \cref{thm:ptwise-convergence}, can be approximated by the square root of the denominator of \cref{eq:pointwise-concergence-formula-semi-classical}. Note that the matrix $K_{\mathrm{full}}$ in \cref{alg:gaussian-process-landmarking} corresponds to the integral kernel in the numerator of \cref{eq:pointwise-concergence-formula-semi-classical} if we choose $\Lambda$ to be the diagonal matrix with $\exp \left[ -V \left( v_i \right)/\epsilon \right]$ at its $i$-th diagonal entry, where $v_i$ is the $i$-th vertex on the triangular mesh; setting $D$ to be the diagonal matrix with the $i$-th row sum of $K_{\mathrm{full}}$ at its $i$-th diagonal entry, the Witten Laplacian $\Delta_{V,\epsilon}$ can then be approximated by $D^{-1/2}K_{\mathrm{full}}D^{-1/2}$ up to a scalar multiplication by the bandwidth $\epsilon$. We would thus expect the first few eigenfunctions corresponding to the smallest eigenvalues of $D^{-1/2}K_{\mathrm{full}}D^{-1/2}$ to concentrate near the critical points of $V$. This can be easily verified with numerical experiments; see \cref{fig:potential-j09} for an example.

\begin{remark}
   The localization effects on eigenfunctions can also be achieved by adding a diagonal matrix, which represents a potential function on the domain of interest, to the kernel matrix; this idea has found applications in biomedical data analysis \cite{CzajaEhler2013} and computer graphics \cite{MRCB2017}. The insight we gained from relating the reweighted kernel to Witten Laplacian reveals that the localization can be equivalently realized through a simple nonlinear procedure. This phenomenon can be explained as follows: while adding a diagonal potential to the kernel provides a direct discretization of the Sch{\"o}dinger operator $L_V=-\Delta+V$, the reweighted kernel \cref{eq:location-prior-kernel} can be thought of as a discretization of the heat kernel of $L_V$ by taking only ``one-hop'' paths in the Feynman-Kac formula for $e^{-tL_V}$. An extensive discussion along this direction is beyond the scope of the current paper; interested readers may find useful the works \cite{SWW2000,SWW2007,Gueneysu2010} and the references therein.
\end{remark}

\begin{figure}[htp]
  \centering
  \includegraphics[width=1.0\textwidth]{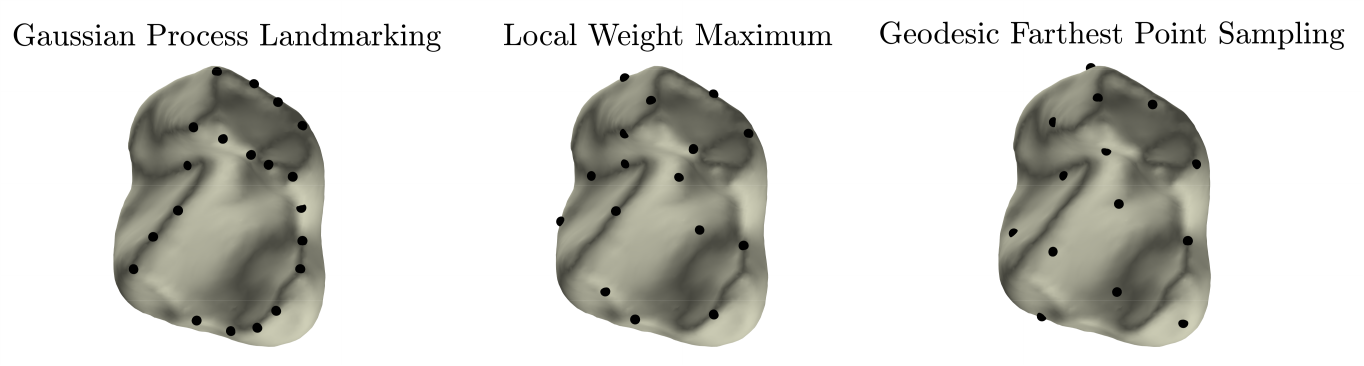}
  \caption{\small An equal number ($20$) of landmark points on the same surface generated using three different strategies. \textbf{Left:} Gaussian process landmarks generated using \cref{alg:gaussian-process-landmarking} with a exponential squared kernel function reweighted by weight function $w$; \textbf{Middle:} The local maxima of the weight function used in the reweighed kernel, which appears semantically less meaningful from the perspective of geometric morphometrics; \textbf{Right:} Points produced by \emph{geodesic farthest point sampling} (see e.g., \cite{Gonzalez1985,MoenningDodgson2003}), a greedy algorithm commonly used for generating uniformly sampled (or approximately space-filling) points on a triangular mesh. Comparing these three sampling approaches, Gaussian process landmarks has the advantage of tending to fill up the manifold while prioritizing the choice of semantically meaningful features for geometric morphometrics.}
  \label{fig:Compare-Landmarks-j01}
\end{figure}

The connection between the reweighted kernel \cref{eq:location-prior-kernel} and the semi-classical analysis of the Witten Laplacian provides insights for the behavior of the Gaussian process landmarks generated from \cref{alg:gaussian-process-landmarking}. Given a Gaussian process $\GP \left( m,K \right)$ defined on a manifold $M$, the eigenfunctions of $K$---properly reweighted by their corresponding eigenvalues--- gives rise to the Karhunen-Lo{\`e}ve basis for $\GP \left( m,K \right)$, with respect to which the sample paths of $\GP \left( m,K \right)$ adopt expansions with i.i.d. standard normal coefficients. If the low-frequency components of these expansions tend to concentrate at certain regions on $M$, when fitting an unknown function in $\GP \left( m,K \right)$ using an active learning procedure, it could be much more efficient if one begins with the inquiry for the function value at those regions of concentration. After information at regions of concentration are collected to some extent, it becomes more beneficial to also incorporate ambient information that resides in the complement of these regions in $M$. Therefore, spreading landmarks on $M$ using a Gaussian process with reweighted kernel balances out the information prioritized by the weight function and the space-filling experimental design strategies. The main reason of not selecting landmarks solely based on the weight function is that the weight function can be too spurious to produce reliable, semantically meaningful landmarks, at least for our application in geometric morphometrics; see \cref{fig:Compare-Landmarks-j01} for an example for such a comparison.

\section{Gaussian Process Landmarking for Automated Geometric Morphometrics}
\label{sec:application-autogm}

This section is divided into three parts. \Cref{sec:unsup-landm-anat} compares Gaussian process landmarks generated from \cref{alg:gaussian-process-landmarking} with ``ground truth'' landmarks manually picked by comparative biologists on a real dataset of anatomical surfaces, demonstrating comparable levels of coverage of biologically significant features. The results presented in this section provide quantitative evidence that Gaussian process landmarks are capable of capturing geometric features encoding important information for comparative biologists on individual anatomical surfaces. \Cref{sec:surf-registr-match} adapts the image feature matching algorithm of \cite{LYPJB2014} for the registration of pairs of anatomical surfaces via matching Gaussian process landmarks computed on each individual surface. We compare the resulting shape correspondence maps with a baseline obtained using previously developed \emph{continuous Procrustes analysis} in \cite{CP13,PNAS2011}; the results suggest that, though Gaussian process landmarks are generated on each individual shape separately, they implicitly encode operationally homologous features that can be compared and contrasted across shapes, and such pairwise comparison results are comparable with those obtained from standard Procrustes shape analysis based on ``ground truth'' observer landmarks. The pairwise shape distances induced from matching landmarks are turned into ordination plots (two-dimensional embeddings of distance matrices, commonly employed for visualizing the ``morphospace'' characterizing shape variances) in \Cref{sec:comp-bio-interp}; it turns out that Gaussian process landmarks lead to a favorable ordination, which is in better agreement with observations made by comparative biologists and paleontologists in existing literature.

We remark that the idea of using landmarks to guide the computation of shape correspondence maps is very natural in the geometry processing community. For instance, \cite{OHG2011} proposes to choose landmarks based on a conditional number quantifying the stability of the matching problem for a pair of shapes, and demonstrated significantly enhanced map quality with very few landmarks specified in this manner; \cite{KLF2011} generates small numbers of feature points with large fractions of semantic correspondences between shapes, and use these points to guide the search for a highly complicated map composed from blending multiple simpler maps. These methods, as well as many other state-of-the-art techniques in geometry processing, are not suitable for general geometric morphometrics applications as they are often fail to capture all anatomical correspondences indispensable for repeatable and reliable analysis.


\subsection{Unsupervised Landmarking on Individual Anatomical Surfaces}
\label{sec:unsup-landm-anat}

Gaussian process landmarks are generated on each anatomical surface individually, regardless of the total size of the shape collection; the landmarks manually selected by human experts (observers), however, may well depend on the information gradually exposed to the human expert as he/she moves through a collection of surfaces. It is thus surprising that the individually generated, ``local'' Gaussian process landmarks bear striking similarity with the observer landmarks selected with certain extent of collection-wise or ``global'' knowledge, as illustrated in \cite[Figure 2]{GPLMK1} through an example fossil molar. This subsection is devoted to a more thorough and quantitative comparison between Gaussian process landmarks and ``ground truth'' observer landmarks placed by human experts.



\begin{figure}[htb]
  \centering
  \includegraphics[width=1.0\textwidth]{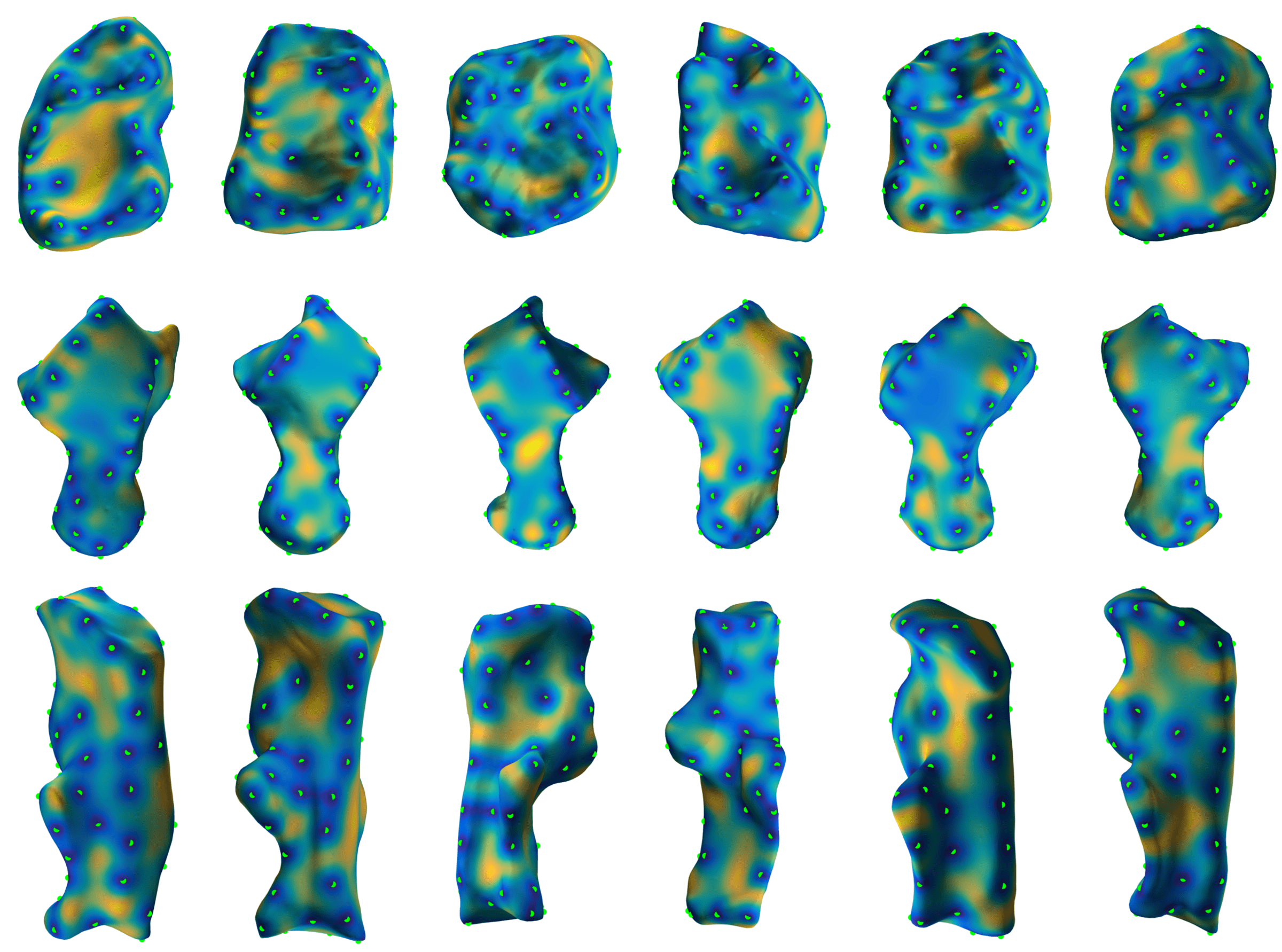}
  \caption{\small Gaussian process landmarks on several different types of anatomical surfaces. All triangular meshes are acquired from $\mu$CT scans. \textbf{Top Row:} Six second mandibular molars of prosimian primates and non-primate close relatives, from a dataset of $116$ molars first published in \cite{PNAS2011}. \textbf{Middle Row:} Six astragalus bones of tarsiers from a dataset of $40$ astragali first published in \cite{Auto3dGM2015}. \textbf{Bottom Row:} Six calcaneus bones of tarsiers from a dataset of $40$ calcanei first published in \cite{Auto3dGM2015}. On all three types of anatomical surfaces, Gaussian process landmarks tend to play the roles of both landmarks and semilandmarks (c.f. \cite[\S2]{ZSS2012}): the curvature-reweighted kernel promotes automatically selecting sharp peaks or saddle points on the anatomical surface; after most of the prominent geometric features---normally recognized as Type 2 landmarks \cite{ZSS2012}---are captured, the uncertainty-based criterion encourages the identification of semilandmark-type points along ridges and grooves.}
  \label{fig:PNAS_subset_short}
\end{figure}

We begin by collecting in \cref{fig:PNAS_subset_short} results obtained by applying the Gaussian Process Landmarking algorithm to several different types of anatomical surfaces, including a subset of the second mandibular molars of primates (first published in \cite{PNAS2011}) and a subset of the astragalus and calcaneus bones of tarsiers first published in \cite{Auto3dGM2015}. It can be recognized from \cref{fig:PNAS_subset_short} that the algorithm is capable of consistently capturing both geometric features (``Type 3 landmarks'' \cite{Roth1993}) and semilandmarks (delineating ridges and grooves).

To quantitatively validate the biological informativeness of Gaussian process landmarks, we use a dataset of second mandibular molars on which the ``observer landmarks,'' or the landmarks manually selected by experienced comparative biologists, are readily available (see \cite{PNAS2011}). We calculate, on each anatomical surface, the median geodesic distance from an observer landmark to its closest Gaussian process landmark. This median geodesic distance can obviously be calculated for any other type of landmarks in place of the Gaussian process landmarks; we shall refer to it as the \emph{median observer-to-automatic landmark distance}, for the sake of simplicity. We compute the median observer-to-automatic landmark distances for a varying number of automatic landmarks, obtaining a curve that encodes the rate with which the median observer-to-automatic landmark distances decay to zero as the number of automatic landmarks increases. Comparing curves obtained from different automatic landmark generation methods then provides us with a way to evaluate how closely each type of automation ``mimics'' the observer landmarks. When we perform such curve comparisons for a large collection of surfaces in a dataset, one can in principle construct statistics (e.g., ``mean'' or ``standard deviation'' in the ``space of curves'') along the lines of \emph{Functional Data Analysis} (FDA, see e.g., \cite{RamsaySilverman2002,RamsaySilverman2005} and the references therein), but we will have to defer such a statistically systematic treatment to future work.

\begin{figure}[htbp]
  \centering
  \includegraphics[width=1\textwidth]{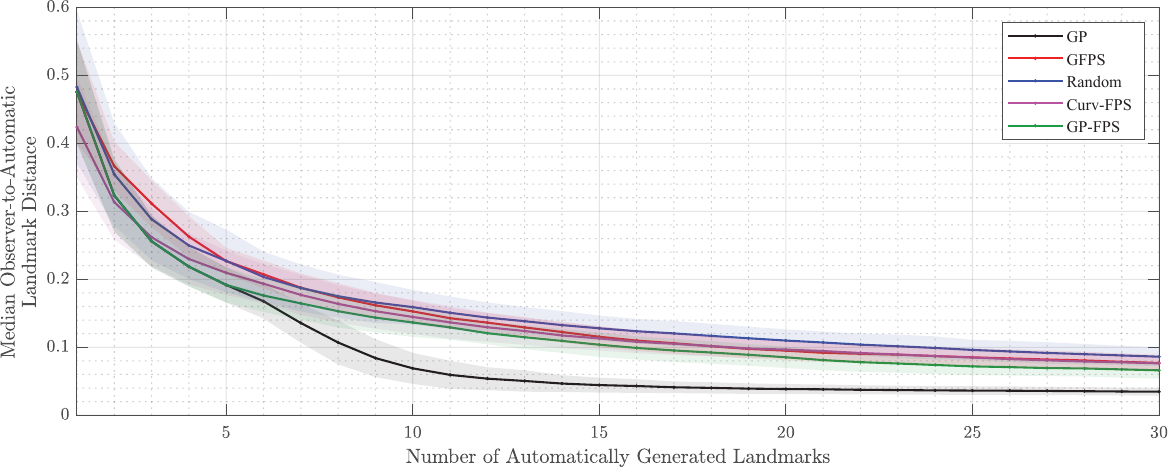}
  \caption{\small Median geodesic distance from an observer landmark to its nearest automatically generated landmark, with respect to different numbers of automatic landmarks, on a collection of $116$ second mandibular molars of prosimian primates and closely related non-primates (see \cite{PNAS2011} for a more detailed description of this dataset and the observer landmark acquisition). Each point on any of the three solid curves is obtained by averaging the $116$ median observer-to-automatic landmark distances over the entire dataset, and the transparent bands represent confidence intervals of one standard deviation. The types of automatic landmarks are Gaussian process landmarks (GP), geodesic farthest point sampling landmarks (GFPS), random landmarks uniformly selected from the vertices of each triangular mesh, and geodesic farthest point sampling landmarks initialized by either critical points of curvature (Curv-FPS) or the first few Gaussian process landmarks (GP-FPS). The random landmarks are only used as a baseline in this experiment.}
  \label{fig:compareCoverage-median}
\end{figure}

Using this strategy, we compare Gaussian process landmarks with landmarks obtained from \emph{Geodesic Farthest Point Sampling} (GFPS) \cite{Gonzalez1985,MoenningDodgson2003}, a popular downsampling technique in automated geometric morphometrics, on a dataset of $116$ second mandibular molars of prosimian primates and closely related non-primates first published in \cite{PNAS2011}. GFPS is known to produce approximately uniformly distributed points on surfaces, with respect to the canonical surface volume measure. In our numerical experiments in this subsection, we choose the first point in GFPS to be the same as the first point obtained by Gaussian process landmarking, to eliminate the effects of random initialization. We further compare with two additional GFPS variants that use a predetermined set of feature points for initialization: Curv-FPS first selects points of critical curvature and then applies geodesic farthest point sampling; we used the setup of~\cite{PNAS2011} for the detection of isolated points of locally critical Gaussian- and mean-curvatures. GP-FPS takes the $k=5$ first Gaussian process landmark points and then proceeds with geodesic farthest point sampling. As a baseline, we also calculate the median observer-to-automatic landmark distances for completely randomly picked vertices on the triangular meshes in this dataset. The results are presented in \cref{fig:compareCoverage-median}, in which each curve is obtained by averaging individual curves over the entire shape collection; confidence intervals of one standard deviation are also plotted in shades of transparency. \cref{fig:compareCoverage-median} suggests that Gaussian process landmarks consistently outperforms GFPS landmarks and the random baseline in terms of coverage of observer landmarks. \cref{fig:compareCoverage-prctiles-mt1}, provided in the supplementary materials, includes additional quantiles of the geodesic distances between observer landmarks and their closest landmarks obtained with each of the automatic methods. The results for the first metatarsals and radii datasets are similar; see \cref{fig:compareCoverage-prctiles-mt1} and \cref{fig:compareCoverage-prctiles-radius} in the supplementary materials.

The comparisons in \cref{fig:compareCoverage-median}, \cref{fig:compareCoverage-prctiles-teeth}, \cref{fig:compareCoverage-prctiles-radius}, and \cref{fig:compareCoverage-prctiles-mt1} are all conducted between the Gaussian process landmarks and the expert observer landmarks manually prepared in \cite{PNAS2011}. These landmarks were treated as ``golden standards'' accepted by domain experts, as show in \cite{PNAS2011} and subsequent domain applications in geometric morphometrics \cite{CP13,GYDMB2017}. It is interesting to investigate into the reproducibility and stability of these observer landmarks --- will Gaussian process landmarks be close to landmarks manually selected by an independent evolutionary anthropologist following the same landmarking protocol (e.g., pre-fixed ordering and/or number of landmarks of each type in the typology summarized in \Cref{sec:geom-morph}, for ensuring the reproducibility of landmark-based geometric morphometrics; see e.g., \cite{Bookstein1991,ZSS2012})? To this end, we collected six additional sets of observer landmarks on the molar teeth in \cref{fig:Q10-ObLmk-vs-GPLmk}, picked by independent human experts following the landmarking protocol detailed in the supplementary materials of \cite{PNAS2011}, and compared them with the observer landmarks in \cite{PNAS2011} and the Gaussian process landmarks. The result is shown in \cref{fig:Comp_UserLmks}. Additional views of this example are provided in the supplementary materials; see \cref{fig:Comp_UserLmks_Supp}. It is clear from these figures that observer landmarks following the same protocol can be robustly reproduced, especially for geometric features and semilandmarks. A thorough investigation of the stability and reproducibility of observer landmarks is beyond the scope of this paper, but the experimental results in \cref{fig:Comp_UserLmks} and \cref{fig:Comp_UserLmks_Supp} suggest that Gaussian process landmarks resemble independent observer landmarks under the same protocol as in \cite{PNAS2011}.

\begin{figure}[htb]
  \centering
  \includegraphics[width=1.0\textwidth]{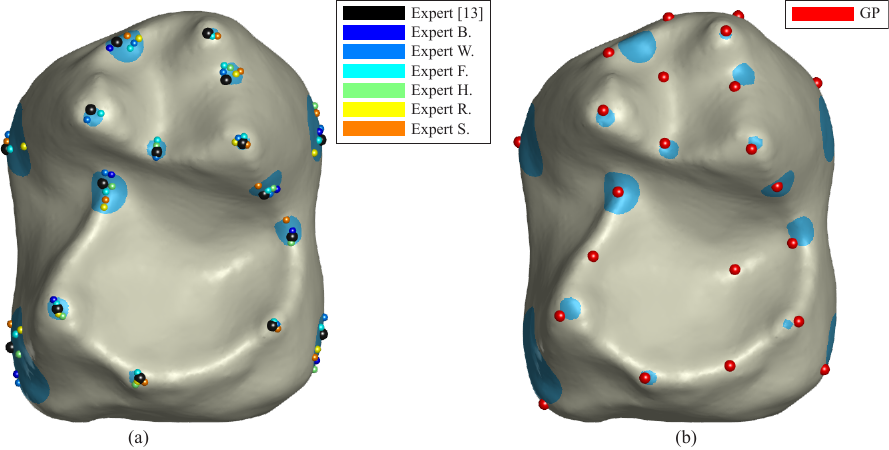}
  \caption{\small (a) Observer landmarks placed by multiple human experts on the fossil molar of \emph{Teilhardina} in \cref{fig:Q10-ObLmk-vs-GPLmk}. Landmarks obtained in \cite{PNAS2011}, shown in black color, were identified manually by an evolutionary anthropologist and used throughout this paper as ground truth. Six additional sets of landmarks are shown in multiple colors; these landmarks were provided by experts guided to manually place landmarks following the same landmarking scheme, so as to capture the same morphological features; see \cref{fig:Comp_UserLmksSupp_Experts} for a visualization of the landmarks provided by each individual expert. Cyan-shaded areas depict the smallest geodesic disk enclosing each group of observer landmarks. (b) Gaussian process landmarks computed with \cref{alg:gaussian-process-landmarking} are shown in red along the cyan disks representing the distribution of human expert landmarks. This further exemplifies the potential of the first few Gaussian process landmarks in identifying biologically-informative feature points.}
  \label{fig:Comp_UserLmks}
\end{figure}


\subsection{Surface Registration via Matching Gaussian Process Landmarks}
\label{sec:surf-registr-match}

We demonstrate in this subsection the benefit of using Gaussian process landmarks for establishing correspondences between pairs of surfaces. In particular, although Gaussian process landmarks are obtained independently on each surface, they turn out to encode geometrically significant features --- shared more often among similar or related shapes --- that are recognized as ``operational homologous'' \cite{Auto3dGM2015} loci by comparative biologists. In order to obtain biologically meaningful correspondences, we use reduced bandwidth parameter in the discrete heat kernel when applying \cref{alg:gaussian-process-landmarking} that is between $3/5$ of the bandwidth parameter used to generate the illustrative \cref{fig:Q10-ObLmk-vs-GPLmk}. Statistical analysis also suggest that these correspondence maps reach comparable explanatory power to observer landmarks placed by human experts, in terms of induced shape distances. The comparison between the morphospaces characterized by these shape distances is deferred to \Cref{sec:comp-bio-interp}.

\subsubsection{Experimental Setup and Methodology}
\label{sec:surf-registr-match-method}

\begin{figure}[t!]
  \centering
  \includegraphics[width=1.0\textwidth]{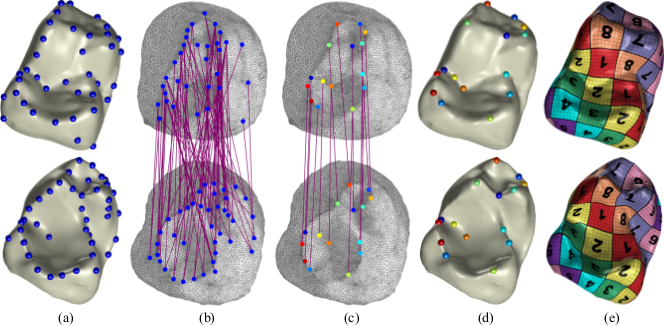}
  \caption{\small The outline of our approach for landmark matching and surface registration. (a) Gaussian process landmarks computed using \cref{alg:gaussian-process-landmarking} on two surfaces; (b) planar parameterizations of the surfaces that minimize~\cref{eq:sym_dirichlet_energy}, overlaid with putative landmark matches (indicated by purple lines); (c) Bounded-Distortion correspondences: a subset of geometrically-consistent matches computed by approximately minimizing \cref{eq:bd_filter_l0}; (d) Pairs of corresponding Gaussian process landmarks found by bounded distortion matching, illustrated by spheres with matching colors; (e) Texture-map visualization of a correspondence map interpolating the landmark correspondences depicted in (d). }
  \label{fig:mathcing_outline}
\end{figure}

\paragraph{Bounded Distortion Gaussian Process Landmark Matching} Let $S_1$, $S_2$ be two-dimensional disk-type surfaces (conformally equivalent to planar disks by the Uniformization Theorem; see e.g., \cite{CP13,LipmanDaubechies2011,LipmanPuenteDaubechies2013}) and let $\xi^{(1)}_1,\ldots,\xi^{(1)}_{L_1}\in S_1$ and $\xi^{(2)}_1,\ldots,\xi^{(2)}_{L_2}\in S_2$ be two sets of Gaussian process landmarks computed using \cref{alg:gaussian-process-landmarking} on $S_1$ and $S_2$, respectively. Note that the algorithm we present in this subsection works equally well for $L_1\neq L_2$, though we choose $L_1=L_2=40$ throughout this paper to simplify the discussion.
Adopting the approach suggested in \cite{LYPJB2014} for feature-based image matching, we devise the following two-step approach for establishing geometrically-consistent matchings between the two sets of Gaussian process landmarks:
\begin{enumerate}[(1)]
\item 
\emph{Parametrization}: For each surface $S_j$ ($j=1,2$) we follow \cite{smith2015,kovalsky2016} to compute an as-isometric-as-possible (AIAP) two-dimensional parametrization, which is a diffeomorphism $\Phi_j:S_j\rightarrow\Omega_j\subset \mathbb{R}^2$ from $S_j$ to a connected planar domain $\Omega_j$ minimizing the (discretization of the) \emph{isometric distortion energy}
\begin{equation} 
\label{eq:sym_dirichlet_energy}
F \left( \phi \right):=\int_{S_j} \left(\left| \nabla\phi \left( x \right)\right|^2 + \left| \nabla\phi^{-1} \left( x \right)\right|^2\right)\dd\mathrm{vol}_M \left( x \right).
\end{equation}
Each landmark $\xi^{(j)}_{\ell}$ is mapped to a unique corresponding point $\zeta^{(j)}_{\ell} = \Phi_j\left( \xi^{(j)}_{\ell} \right)\in \Omega_j\subset \mathbb{R}^2$, where
\begin{equation*}
  \Phi_j:=\argmin_{\phi:S_j\rightarrow\mathbb{R}^2} F \left( \phi \right).
\end{equation*}
Since both ``left-handed'' and ``right-handed'' shapes exist in our dataset (as illustrated in \Cref{fig:PNAS_subset_short}), we attempt two parametrizations --- with or without reflection --- for each surface, and only keep the orientation that leads to a smaller Procrustes score (see \eqref{eq:proc_distance} below). A byproduct of this implementation detail is a consistent re-arrangement of the surface orientations.

\item \emph{Bounded Distortion Matching}: Following \cite{LYPJB2014}, we search within the set of planar diffeomorphisms between $\Omega_1$ and $\Omega_2$ with \emph{conformal distortion} \cite{Lipman2012} bounded by a pre-fixed constant $K\geq 1$. This algorithm strives to find a maximal subset of geometrically-consistent correspondences within an initial set of candidate matches. In the extreme case of $K=1$, the search is constrained within the set of strictly angle-preserving (conformal) maps for the continuous isometric distortion energy and the set of planar rigid transformations for the discretized isometric distortion energy; we select $K=1.5$ in this experiment to slightly enlarge the search space of candidate maps.

As input to this matching algorithm, for each Gaussian process landmark $\zeta_{\ell}^{\left( 1 \right)}$ on $S_1$, we choose $T\geq 2$ Gaussian process landmarks $\zeta^{(2)}_{\ell\to 1},\ldots,\zeta^{(2)}_{\ell\to T}$ as initial putative candidate matches from the Gaussian process landmarks $\left\{\zeta_j^{\left( 2 \right)}\mid 1\leq j\leq L_2\right\}$ on $S_2$; the algorithm then searches for the bounded distortion map $\Psi:\Omega_1\rightarrow\Omega_2$ that approximately minimizes the mismatch count
\begin{equation}
\label{eq:bd_filter_l0}
\sum_{\ell=1}^{L_1}\sum_{k=1}^T \left\| \Psi\left( \zeta^{(1)}_{\ell} \right) - \zeta^{(2)}_{\ell\to k} \right\|^0,
\end{equation}
where, following the notations of \cite{LYPJB2014}, $\left\|\cdot\right\|^0$ denotes the mixed $\left(2,0\right)$-norm
\begin{equation*}
  \left\| \Psi\left( \zeta^{(1)}_{\ell} \right) - \zeta^{(2)}_{\ell\to k} \right\|^0=
  \begin{cases}
    1 & \textrm{if $\Psi\left( \zeta^{(1)}_{\ell} \right) \neq \zeta^{(2)}_{\ell\to k}$,}\\
    0 & \textrm{otherwise.}
  \end{cases}
\end{equation*}
In practice, we follow \cite{LYPJB2014} and approximate the mixed $\left(2,0\right)$-norm with a mixed $\left(2,p\right)$-norm with $p$ decreasing during the optimization.
The initial candidate matches in our experiments are generated by comparing the \emph{Wave Kernel Signature} (WKS) \cite{ASC2011} of the Gaussian process landmarks --- the $T$ Gaussian process landmarks on $S_2$ with most similar WKS's (measured in Euclidean distances) with that of each $\zeta_{\ell}^{\left( 1 \right)}\in S_1$ are selected as $\left\{\zeta^{(2)}_{\ell\to t}\mid 1\leq t\leq T\right\}$. Ideally, ``incorrect'' initial matches, which potentially lead to large conformal distortions, will be filtered out by minimizing \cref{eq:bd_filter_l0} under the bounded conformal distortion constraint. Our implementation uses $T=2$; see \cref{fig:mathcing_outline}(b) for an illustrative example. 
\end{enumerate}

The pairwise registration algorithm outputs a subset of $1\leq L\leq \min\left\{L_1,L_2\right\}$ one-to-one correspondences $\tilde{\xi}^{(1)}_{\ell} \leftrightarrow \tilde{\xi}^{(2)}_{\ell}$, $\ell=1,\ldots,L$, from the initial $L_1 T$ candidate Gaussian process landmark matches; the number of matched landmarks, $L$, is automatically determined by the final bounded distortion map and could be vary between different pairs of surfaces. See \cref{fig:mathcing_outline}(c)(d) for an illustrative example. A final step in this algorithm pipeline ``interpolates'' the $L$ pairs of matched Gaussian process landmarks to obtain a diffeomorphism between the surfaces $S_1$ and $S_2$; see \cref{fig:mathcing_outline}(d) for an illustrative example. For this purpose, we use the technique developed in \cite{smith2015,kovalsky2016} which computes a map $\tilde{\Psi}:\Omega_1 \rightarrow \Omega_2$ that minimizes the isometric distortion energy \cref{eq:sym_dirichlet_energy} subject to the $L$ linear equality constraints $\tilde{\Psi}\left(\tilde{\zeta}^{(1)}_{\ell}\right) = \tilde{\zeta}^{(2)}_{\ell}$ ($1\leq \ell \leq L$) representing the ``sparse'' correspondences between Gaussian process landmarks. The composition $f_{12}={\Phi_2}^{-1}\circ\tilde{\Psi}\circ\Phi_1$ finally produces the desired map $f:S_1\to S_2$ that approximately minimizes the mismatch count~\cref{eq:bd_filter_l0}. This last interpolation step is indispensable for the purpose of interpretability (producing full-surface registrations for visual comparisons) as well as evaluation (inducing shape Procrustes distances for ordination, see \Cref{sec:comp-bio-interp}). \cref{fig:mathcing_outline} outlines the complete workflow of landmark matching and surface registration for a pair of molars surfaces from \cite{PNAS2011}; a few examples are shown in \cref{fig:maps_examples}.

\paragraph{Evaluation Metrics and the Baseline} We demonstrate the rich taxonomic information captured by the maps output from the proposed bounded distortion Gaussian process landmark matching (GP-BD) algorithm, by comparing the shape \emph{Procrustes distances} induced by maps computed from GP-BD with those obtained by the same feature matching methodology but alternative choices of landmarking schemes. For any correspondence map $f:S_1\to S_2$, we define the \emph{Procrustes distance induced by $f$} as
\begin{equation} \label{eq:proc_distance}
  d_\textrm{P} \left( f \right)=\left(\min_{R\in \mathbb{E}_3}\int_{S_1} \left\| f \left( x \right)-R\left(x\right) \right\|^2 \dd\mathrm{vol}_{S_1}\left( x \right)\right)^{\frac{1}{2}},
\end{equation}
where $\mathbb{E}_3$ stands for the rigid motion group in $\mathbb{R}^3$. The distance $d_\text{P} \left( f \right)$ measures the spatial registration error induced by the map $f$ between the two surfaces. To ensure that the Procrustes distances induced by the various methods listed in \cref{table:experiment_abbv} is comparable, we normalize each surface to have unit surface area. When the map $f$ is the continuous Procrustes map produced from the algorithm presented in \cite{CP13}, $d_{\textrm{P}}$ gives exactly the \emph{continuous Procrustes distance} between $S_1$ and $S_2$.

The alternatives considered in this section include two other different types of Gaussian process landmarks (GP$_\textrm{nW}$-BD and GP$_\textrm{Euc}$-BD) with alternative kernel functions (different from \cref{eq:location-prior-kernel-discrete}), two different strategies of utilizing the ``ground truth'' user-placed observer landmarks (GT-BD and GT$^2$-BD) as ``oracles'' representing the ``ideal'' landmarks for feature matching, as well as the baseline \emph{continuous Procrustes maps} (CPM) reported in \cite{CP13,PNAS2011}; \cref{table:experiment_abbv} provides a summary of these variants. Specifically, GP$_\textrm{Euc}$-BD and GP$_\textrm{nW}$-BD are based on Gaussian processes with the standard squared exponential kernel \cref{eq:sq-exp-kernel-submfld} and the trivially weighted kernel (setting $w\equiv 1$ in \cref{eq:location-prior-kernel}), respectively. GT-BD simply replaces the GP landmarks with ``ground truth'' (GT) landmarks, which then follows through with candidate selection via WKS and bounded distortion maps based feature matching filtering (BD-filtering); GT$^2$-BD also takes ground truth landmarks as inputs, but skips the candidate selection by setting the ground truth correspondences between observer landmarks as candidate matches, though BD-filtering still applies and prunes out potential ``geometrically incompatible'' correspondences leading to large conformal distortions. A detailed description of observer landmark acquisition can be found in \cite{PNAS2011}.

The shape distances will be compared in several different ways. We will first compare cumulative distributions of the pairwise shape distance values, followed by two statistical tests addressing (i) the correlation between each automatic shape distance and the observer-determined landmarks Procrustes distance (ODLP), and (ii) the capability of each shape distance at distinguishing taxonomic groups. A qualitative but more intuitive comparison of the \emph{morphospaces} \cite{MitteroeckerHuttegger2009} characterized by these shape distances will be deferred to supplementary materials \Cref{sec:comp-bio-interp}, in the form of ordination plots (two dimensional embedding of the shape distance matrices as visual representations for shape variances across species groups). The discussion in \Cref{sec:comp-bio-interp} is however oriented slightly more towards readers with some background in comparative biology.

\begin{table}[t]
  \footnotesize
\centering
\begin{tabular}{|l|p{10cm}|}
\hline
Abbreviation & Description of the pairwise surface registration method \\
\hline
GP-BD & BD-filtering for GP landmarks computed with \cref{alg:gaussian-process-landmarking} \\
\hline
  GP$_\textrm{Euc}$-BD & BD-filtering for GP landmarks with the standard Euclidean heat kernel \cref{eq:sq-exp-kernel-submfld} \\
\hline
  GP$_\textrm{nW}$-BD & BD-filtering for GP landmarks with non-weighted kernel ($w\equiv 1$ in \cref{eq:location-prior-kernel-discrete}) \\
\hline
  GT-BD &  BD-filtering for GT landmarks \\
\hline
  GT$^2$-BD & BD-filtering for GT landmarks, also using the ground truth correspondences as candidate initial matches\\
\hline
  CPM & Continuous Procrustes Maps (CPM) computed using the method of \cite{PNAS2011,CP13} \\ 
\hline
\end{tabular}
\caption{\small The pairwise surface registration methods compared in \Cref{sec:surf-registr-match} and \Cref{sec:comp-bio-interp}. BD-filtering stands for \emph{bounded distortion maps based feature matching filtering}; GP stands for ``Gaussian process''; GT stands for ``Ground Truth,'' i.e., those landmarks placed by experienced comparative biologists.}
\label{table:experiment_abbv}
\end{table}

\subsubsection{Comparison Results}
\label{sect:surface_maps}

\begin{figure}[t!]
  \centering
  \includegraphics[width=1\textwidth]{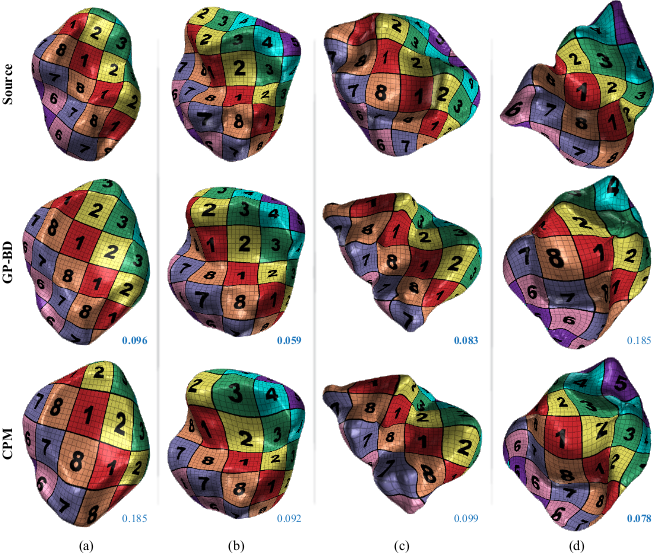}
  \caption{\small Correspondence maps and induced Procrustes distances: texture overlaid on the Source surface (top row) is mapped by $f:S_1\to S_2$ so as to visualize correspondence. The second row (GP-BD) shows maps induced by correspondences established between Gaussian process landmarks computed with \cref{alg:gaussian-process-landmarking}. The bottom row (CPM) compares the baseline continuous procrustes maps. GP-BD outperforms CPM in examples (a)-(c), wherein the map it induces establishes a better correspondence between morphological features of the surfaces; example (d) shows a case in which GP-BD is outperformed by CPM. The inset values are the the Procrustes distance $d_\textrm{P}$ associated with each map.}
  \label{fig:maps_examples}
\end{figure}

We compared the pairwise surface registration methods listed in \cref{table:experiment_abbv} on three different dataset of anatomical surfaces: (i) $116$ second mandibular molars of prosimian primates and closely related non-primates; (ii) $57$ proximal first metatarsals of prosimian primates, New and Old World monkeys; (iii) $45$ distal radii of apes and humans. Detailed descriptions of all $3$ datasets can be found in \cite{PNAS2011}. We computed correspondences between each pair of surfaces within this dataset, totaling in over $13,000$ correspondence maps for dataset (i), over $1,500$ for dataset (ii), and nearly $1,000$ for dataset (iii). \cref{fig:maps_examples} shows example pairs of surfaces, visualizing correspondence maps induced by GP-BD correspondences as well as the baseline CPM maps. These examples illustrate typical differences between maps computed from GP-BD and CPM along with their Procrustes distances; GP-BD maps often offer an improvement over CPM in terms of both visual quality and their ability to relate biologically meaningful and operationally homologous regions.

\begin{figure}[t!]
  \centering
  \includegraphics[width=.85\textwidth]{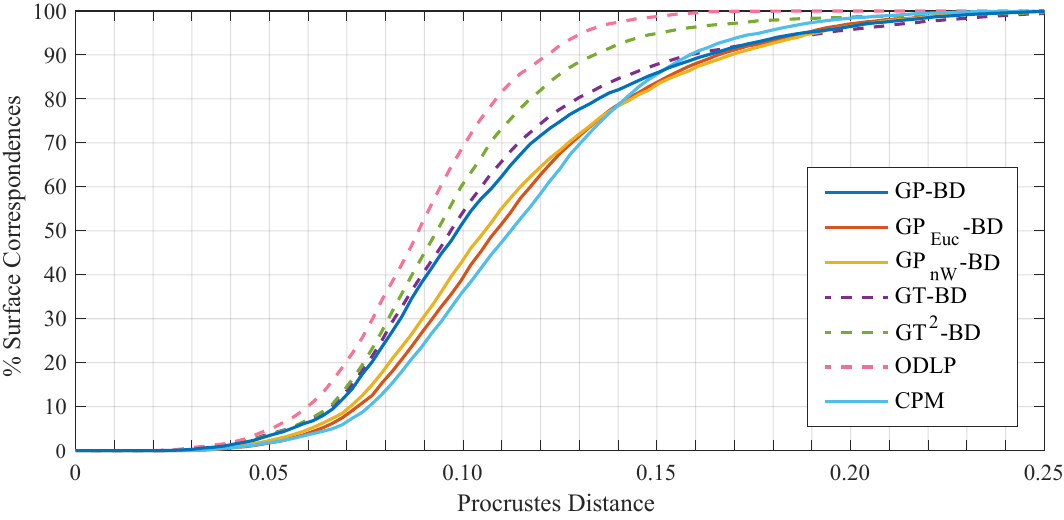}
  \caption{\small Comparing GP-BD to the alternative methods listed in \cref{table:experiment_abbv}. Each curve represents the proportion of correspondences, among a collection of 116 surfaces, for which the Procrustes distance $d_\textrm{P}$ falls below a given threshold. The cumulative distribution of GP-BD distances most closely resembles that of GT-BD which utilizes the ground truth observer landmarks. It falls short of GT$^2$-BD and the Observer-Determined Landmarks Procrustes (ODLP), both of which rely additionally on the ``ground truth'' correspondences between observer landmarks. In terms of producing more relatively smaller distances between the shape pairs (which is favorable by the correlation between smaller distances and enhanced morphometric interpretability reported in \cite{GYDMB2017}; also suggested by the cumulative distribution curves of ODLP, GT$^2$-BD, and GT-BD), GP-BD outperforms the baseline continuous Procrustes maps (CPM), as well as variants of Gaussian process landmarks with alternative kernel constructions, GP$_\textrm{nW}$-BD (Gaussian process landmarks with a non-weighted kernel) and GP$_\textrm{Euc}$-BD (Gaussian process landmarks with a standard Euclidean heat kernel).}
  \label{fig:error_plot}
\end{figure}

\paragraph{Comparing Cumulative Distributions of Distance Values} We first provide a crude comparison across the various methods listed in \cref{table:experiment_abbv}. \cref{fig:error_plot} plots the cumulative distributions of Procrustes distances induced by each type of pairwise registration method in \cref{table:experiment_abbv}, which serves as a direct comparison of the proportions of pairwise correspondences for which the Procrustes distance $d_\textrm{P}$ \cref{eq:proc_distance} is below a given threshold. The figure further includes a curve for the Observer-Determined Landmarks Procrustes (ODLP) distances, computed using the standard Procrustes analysis between sets of corresponding landmarks manually placed by human experts \cite{PNAS2011}. Recall from \Cref{sec:surf-registr-match-method} that the surface areas are normalized so as to ensure that Procrustes distances induced by different shape correspondences are comparable.

Noticeably, the cumulative distribution of GP-BD distances most closely resemble that of GT-BD, obtained by replacing the Gaussian process landmarks with the ground-truth observer landmarks (but otherwise using exactly the same algorithm involving WKS and BD-filtering). Also, GP-BD outperforms GP$_\textrm{nW}$-BD (Gaussian process with a trivially weighted kernel) and GP$_\textrm{Euc}$-BD (Gaussian process with the standard Euclidean heat kernel). This comparison suggests that the Gaussian process landmarks computed with \cref{alg:gaussian-process-landmarking} provide a good proxy for geometrically significant features needed to determine meaningful correspondences between surfaces.


Expectedly, GP-BD falls short compared to distances that rely on both the ground-truth observer landmarks and their true correspondences (i.e., GT$^2$-BD and ODLP). Comparing GP-BD to the baseline continuous Procrustes maps (CPM) in terms of distance distributions is equivocal as the two curves in \cref{fig:error_plot} cross each other when the Procrustes distance threshold is about $0.15$. Nonetheless, comparing GP-BD to CPM indicates that the former produces more pairwise correspondences with shape distances less than $0.1$ (at which the vertical gap between the two curves reaches its maximum). This, along with the correlation reported in \cite{GYDMB2017} between smaller continuous Procrustes (cP) distances and better morphometric interpretability of the associated maps, implies that Gaussian process landmarks potentially lead to more stable and interpretable comparative biological analysis if combined with other globally transitive geometric morphometric methods.

 
\paragraph{Statistical Tests Comparing the Distance Matrices} We now provide a more thorough comparison, using nonparametric statistical tests, for the shape distance matrices induced by the various methods listed in \cref{table:experiment_abbv}. Treating the OPLD (Observer-Determined Landmarks Procrustes distance) matrix as a proxy for the ``ground truth'' accepted among geometric morphometricians (c.f. \cite{PNAS2011}), we first compare the correlation between each automatically computed shape distance matrix with ODLP using the \emph{Mantel correlation test} \cite{Mantel1967}; the explanatory power of the shape distances (in terms of the ability to distinguish taxonomic groups) are then compared using a variant of the multivariate analysis of variance (MANOVA) for distance matrices.


Due to the dependence of entries in a distance matrix (e.g., constrained by the triangular inequality), assessments of correlation between distance matrices, or between a distance matrix and a (continuous or categorical) variable, often involves repeatedly permuting the rows and columns of the distance matrix (see e.g., \cite{Pesarin2001,NicholsHolmes2002,Good2004} and the references therein). 
\cref{table:mantel} demonstrates the results of Mantel correlation analysis \cite{Mantel1967} between the distance matrices computed in this paper against the ODLP distance matrix. In all mantel tests in this paper, we set the number of permutations to $9999$ since it is recommended in \cite{Anderson2001} that at least $5000$ permutations should be done for tests with an $\alpha$-level of $0.01$. GT-BD and GT$^2$-BD correlate best with ODLP, which is as expected since they both rely directly on the same set of observer-determined landmarks used for computing the ODLP distances; their high correlation with ODLP justifies the use of bounded distortion maps for capturing biologically meaningful and corresponding geometric features. GP-BD outperforms CPM, which is consistent with our observation in \cref{fig:error_plot} that a majority (over $80\%$) of the distance values computed from GP-BD is more similar in distribution to the distances computed with ground truth landmarks. The advantageously higher correlation of GP-BD over GP$_{\textrm{Euc}}$ and GP$_{\textrm{nW}}$ illustrates the importance of reweighting in the kernel construction.


\begin{table}[htb]
\centering
\begin{tabular}{|l|cc|c|ccc|}
\hline
& GT & GT$^2$ & CPM & GP & GP$_{\textrm{Euc}}$ & GP$_{\textrm{nW}}$ \\
\hline
  Molars & $\mathbf{0.7042}$ & $\mathbf{0.7563}$ & $0.6647$  & $0.6870$ & $0.6135$ & $0.6257$ \\
  \hline
  First Metatarsals & $\mathbf{0.7371}$ & $\mathbf{0.8326}$ & $0.4887$ & $0.7318$ & $0.6607$ & $0.7117$ \\\hline
  Radii & $\mathbf{0.3273}$ & $\mathbf{0.4909}$ & $0.1775$ & $0.3231$ & $0.2510$ & $0.2746$ \\
\hline
\end{tabular}
\caption{\small Correlation coefficients output from Mantel correlation analysis for the various distances computed in this paper versus ODLP distances, for three different datasets. See \cref{table:experiment_abbv} for the list of abbreviations; note that we omitted ``-BD'' for the sake of space. The relatively high correlations of GT-BD and GT$^2$-BD versus ODLP is not surprising due to their direct dependence on the observer-determined landmarks; the first $2$ columns can thus be viewed as performance upper bounds when ``oracle landmark correspondences'' are provided. For all three datasets, the distance matrix computed from GP-BD correlates better with ODLP than CPM, as well as almost all of its variants, which is consistent with the CDF plot \cref{fig:error_plot} and the ordination plots in \cref{fig:species-ordination-plots} and speaks of the advantages of matching anatomical surfaces with Gaussian process landmarks. The $P$-values of all results are $<0.01$.}
\label{table:mantel}
\end{table}

In addition, we perform \emph{Permutational Multivariate Analysis of Variance} (PERMANOVA) \cite{Anderson2001} for the distance matrices computed in this paper, with the taxonomic groups shown in the ordination plots (\cref{fig:species-ordination-plots}, see detailed explanations in \Cref{sec:comp-bio-interp}) as the treatment effects for the molar dataset; for the first metatarsal and radius datasets, we use family and species groups as treatment effects, respectively. The results are presented in \cref{table:permanova}. The purpose of this test is to quantitatively compare the power of differentiating taxonomic groups for these distance matrices. The pseudo $F$-statistics in PERMANOVA is a properly normalized ratio between among-group and within-group sum of squared distances. The statistical significance is then calculated from the fraction---among sufficiently many shuffles of the rows and columns of the distance matrix---of the permutation instances that produce a higher pseudo $F$-ratio. While all distance matrices demonstrate statistically significant pseudo $F$-ratios in this test, the GT-BD distance matrix leads the board of pseudo $F$-ratios, indicating its superior ability of separating taxonomic groups; the GP-BD distance matrix and its two variants consistently outperform CPM in terms of pseudo $F$-ratios, verifying again the improved quality of pairwise anatomical surface registrations as well as the induced shape Procrustes distances.

\begin{table}[htb]
  \centering
  \begin{tabular}{|l|ccc|c|ccc|}
\hline
 & ODLP  & GT & GT$^2$ & CPM & GP & GP$_{\textrm{Euc}}$ & GP$_\textrm{nW}$\\
\hline
    Molars & $12.26$ & $\mathbf{16.70}$ & $13.90$ & $9.42$ & $\mathbf{16.26}$ & $11.15$ & $12.99$ \\
    \hline
    First Metatarsals & $20.30$ & $\mathbf{29.87}$ & $18.12$ & $6.38$ & $\mathbf{32.85}$ & $19.17$ & $21.82$ \\\hline
    Radii & $\mathbf{9.19}$ & $\mathbf{11.89}$ & $9.03$ & $5.59$ & $9.09$ & $6.35$ & $8.04$ \\
  \hline
\end{tabular}
\caption{\small Pseudo $F$-ratios output from PERMANOVA for various distance matrices computed in this paper for three different datasets. See \cref{table:experiment_abbv} for the list of abbreviations; note that we omitted ``-BD'' for the sake of space. The number of groups for the molar dataset is $30$, equaling to the number of polygonal regions in the ordination plot in \cref{fig:species-ordination-plots}; the numbers of groups for first metatarsals and radius are $8$ (families) and $4$ (genuses), respectively. The first column refers to observer-determined landmarks Procrustes (ODLP) distances calculated in \cite{PNAS2011} as a baseline; the first $3$ columns can thus be viewed as performance upper bounds when ``oracle landmark correspondences'' are provided. For all three datasets, the pseudo-$F$ ratio of the GP-BD distance matrix is better or at least comparable to the oracles. All distance matrices are statistically significant in terms of their powers of separating species groups: the $P$-values of all results are $<0.01$.}
\label{table:permanova}
\end{table}

\section{Discussions}
\label{sec:conclusions}

In this paper, we apply the uncertainty-based landmark generating algorithm proposed in \cite{GPLMK1} to three-dimensional geometric morphometrics. The algorithmically produced landmarks resemble biologists' landmarks selected with expert knowledge, providing adequate coverage for both geometric features and semantically ``uncertain'' regions on the anatomical surfaces of practical interest. We tested the applicability of this landmarking procedure for various tasks on real datasets.

The Gaussian process landmarking algorithm presented in this paper takes one anatomical surface as input at one time, which is not exactly consistent with the methodology of geometric morphometricians. In fact, these biologists do not simply place landmark on individual surface; rather, they are trained to take into consideration an entire collection of shapes, so as to consistently place landmarks that are in joint correspondence. In standard practice, landmarking a new anatomical surface typically involve repeated comparisons with all the other surfaces in the collection, and the already placed landmarks on a surface are still subject to change upon future knowledge acquired from landmarking more surfaces. It would be highly interesting to adapt our algorithm to accommodate for this type of group-wise comparison strategies as well.


Though geometric approaches in general lack the ability to recognize Type 1 landmarks (recall Bookstein's typology reviewed in \Cref{sec:geom-morph}), we think that the similarities between Gaussian process landmarks and user selected Type 2 and Type 3 landmarks call into question user-based landmarks. In particular, we found the ordination plots to be as successful, and in some ways superior, to the user-based landmarks in reflecting previous ideas about shape affinities. While $F$-ratios were highest in the user-determined sample, we note that it is highly likely that, for any underdetermined landmarks (e.g., the Type 1 landmarks, Type 3 landmarks, or when trying to place Type 2 landmarks on eroded features like blunted cusp tips) biologists will tend to minimize variance within species. In other words, if there is not enough geometry to allow consistent placement of a point, users are likely to unconsciously choose a positioning that visually maximizes similarity to other members of the species. Given the likely bias towards minimizing within group error by biologists during landmarking, it is quite remarkable that the Gaussian process approach comes so close to the ``ground truth'', and exceeds by so much the other automated methods in terms of $F$-ratio.  In user-placed landmarks, there is also the question of how many and which landmarks were chosen, not to mention whether different users are as accurate in placing the same landmarks. Finally, there is the limitation that any traditional landmark used must be present in every specimen of the sample. Given the challenges with the traditional user based approach, and the demonstrated ability of Gaussian process to emulate qualitatively and quantitatively Type 2 and Type 3 landmarks, we believe the Gaussian process landmarking algorithm has great potential to be an alternative, automated approach for such landmarks.

\section*{Software}
\verb|MATLAB| code for the surface registration algorithm is publicly available at \url{https://github.com/shaharkov/GPLmkBDMatch}.

\section*{Acknowledgments}
The authors thank Peng Chen, Chen-Yun Lin, Shaobo Han, Sayan Mukherjee, Rob Ravier, and Shan Shan for inspirational discussions; Ethan Fulwood, Bernadette Perchalski, Julia Winchester, Arianna Harrington, Robert Ravier, and Shan Shan for assistance with collecting observer landmarks; the anonymous reviewers for many constructive feedback.

\appendix

\section{Proof of \cref{thm:ptwise-convergence}}
\label{sec:proof-theorem-ptws-convergence}

\begin{proof}[Proof of \cref{thm:ptwise-convergence}]
  Recall from \cite{Singer2006ConvergenceRate} (or \cite[Lemma 8]{CoifmanLafon2006}, \cite[Lemma B.10]{SingerWu2012VDM}) that
  \begin{equation}
    \label{eq:basic-expansion}
    \begin{aligned}
      \frac{1}{\left( 2\pi\epsilon \right)^{\frac{d}{2}}}&\int_M\exp \left( -\frac{\left\| x-y \right\|^2}{2\epsilon} \right)f \left( y \right)\dd\mathrm{Vol}\left( y \right)\\
      &=f \left( x \right)+\frac{\epsilon}{2} \left[ E \left( x \right)f \left( x \right)+\Delta f \left( x \right) \right]+O \left( \epsilon^{\frac{3}{2}} \right)\qquad\textrm{as $\epsilon\rightarrow 0$}
    \end{aligned}
  \end{equation}
where $E$ is a scalar function of the curvature of $M$ at $x\in M$. Thus as $\epsilon\rightarrow0$
\begin{equation}
  \label{eq:inner-integral}
  \begin{aligned}
    \exp &\left[ -V \left( z \right) \right]\int_M\exp \left( -\frac{\left\| z-y \right\|^2}{2\epsilon} \right) f\left( y \right)\dd\mathrm{Vol}\left( y \right)\\
    &=\left( 2\pi\epsilon \right)^{\frac{d}{2}} \exp \left[ -V \left( z \right) \right]\left\{ f \left( z \right)+\frac{\epsilon}{2} \left[ E \left( z \right)f \left( z \right)+\Delta f \left( z \right) \right]+O \left( \epsilon^{\frac{3}{2}} \right) \right\}.
  \end{aligned}
\end{equation}
Replacing the function $f$ in \cref{eq:basic-expansion} with the right hand side of \cref{eq:inner-integral}, we have
\begin{align*}
    \int_M&\!\int_M\exp \left( -\frac{\left\| x-z \right\|^2}{2\epsilon} \right)\exp \left[ -V \left( z \right) \right]\exp \left( -\frac{\left\| z-y \right\|^2}{2\epsilon} \right)f \left( y \right)\dd\mathrm{Vol}\left( z \right)\dd\mathrm{Vol}\left( y \right)\\
    &= \left( 2\pi\epsilon \right)^d \Bigg\{ \exp \left[ -V \left( x \right) \right]\left\{ f \left( x \right)+\frac{\epsilon}{2} \left[ E \left( x \right)f \left( x \right)+\Delta f \left( x \right) \right]+O \left( \epsilon^{\frac{3}{2}} \right) \right\}\\
    &\quad+\frac{\epsilon}{2} E \left( x \right)\exp \left[ -V \left( x \right) \right]\left\{ f \left( x \right)+\frac{\epsilon}{2} \left[ E \left( x \right)f \left( x \right)+\Delta f \left( x \right) \right]+O \left( \epsilon^{\frac{3}{2}} \right) \right\}\\
    &\quad+\frac{\epsilon}{2}\Delta \left( \exp \left[ -V \left( x \right) \right]\left\{ f \left( x \right)+\frac{\epsilon}{2} \left[ E \left( x \right)f \left( x \right)+\Delta f \left( x \right) \right]+O \left( \epsilon^{\frac{3}{2}} \right) \right\} \right)+ O \left( \epsilon^{\frac{3}{2}} \right) \Bigg\}\\
    &= \left( 2\pi\epsilon \right)^d \Bigg\{ f \left( x \right)\exp \left[ -V \left( x \right) \right]+\frac{\epsilon}{2}\Big\{\left[ E \left( x \right)f \left( x \right)+\Delta f \left( x \right) \right]\exp \left[ -V \left( x \right) \right]\\
    &\quad+ f \left( x \right)E \left( x \right)\exp \left[ -V \left( x \right) \right]+\Delta \Big( f \left( x \right) \exp \left[ -V \left( x \right) \right] \Big)\Big\} + O \left( \epsilon^{\frac{3}{2}} \right)\Bigg\}\qquad \textrm{as $\epsilon\rightarrow 0$.}
\end{align*}

Plugging the right hand side above into the numerator of the left hand side of \cref{eq:pointwise-concergence-formula}, and replacing the denominator with the same expression as the numerator except for setting $f\equiv 1$ in \cref{eq:pointwise-concergence-formula}, we obtain, as $\epsilon\rightarrow 0$,
\begin{align*}
    &\frac{\displaystyle\int_M\!\int_M\exp \left( -\frac{\left\| x-z \right\|^2}{2\epsilon} \right)\exp \left[ -V \left( z \right) \right]\exp \left( -\frac{\left\| z-y \right\|^2}{2\epsilon} \right)f \left( y \right)\dd\mathrm{Vol}\left( z \right)\dd\mathrm{Vol}\left( y \right)}{\displaystyle\int_M\!\int_M\exp \left( -\frac{\left\| x-z \right\|^2}{2\epsilon} \right)\exp \left[ -V \left( z \right) \right]\exp \left( -\frac{\left\| z-y \right\|^2}{2\epsilon} \right)\dd\mathrm{Vol}\left( z \right)\dd\mathrm{Vol}\left( y \right)}\\
    &=\Bigg\{ f \left( x \right) + \frac{\epsilon}{2}\left[ 2 E \left( x \right)f \left( x \right)+\Delta f \left( x \right)+\exp \left[ V \left( x \right) \right]\Delta \Big( f \left( x \right)\exp \left[ -V \left( x \right) \right] \Big) \right]\\
    &\qquad\qquad+O \left( \epsilon^{\frac{3}{2}} \right) \Bigg\}\times\Bigg\{1-\frac{\epsilon}{2} \Big( 2 E \left( x \right)+\exp \left[ V \left( x \right)\right] \Delta \exp \left[ -V \left( x \right) \right] \Big) +O \left( \epsilon^{\frac{3}{2}} \right) \Bigg\}\\
    &=f \left( x \right)+\frac{\epsilon}{2}\Delta f \left( x \right)+\frac{\epsilon}{2}\exp \left[ V \left( x \right) \right]\Delta \left( f \left( x \right)\exp \left[ -V \left( x \right) \right] \right)\\
    &\qquad\qquad\qquad\qquad\quad-\frac{\epsilon}{2}f \left( x \right)\exp \left[ V \left( x \right) \right]\Delta\exp \left[ -V \left( x \right) \right]+O \left( \epsilon^{\frac{3}{2}} \right)\\
    &=f \left( x \right)+\epsilon \Big(\Delta f \left( x \right) - \nabla f \left( x \right)\cdot \nabla V \left( x \right)\Big)+O \left( \epsilon^{\frac{3}{2}} \right).
\end{align*}
This completes the proof.
\end{proof}

\section{Numerical Experiments on Additional Datasets}
\label{sec:numer-exper-addit}

\begin{figure}[htpb]
  \centering
  \includegraphics[width=.85\textwidth]{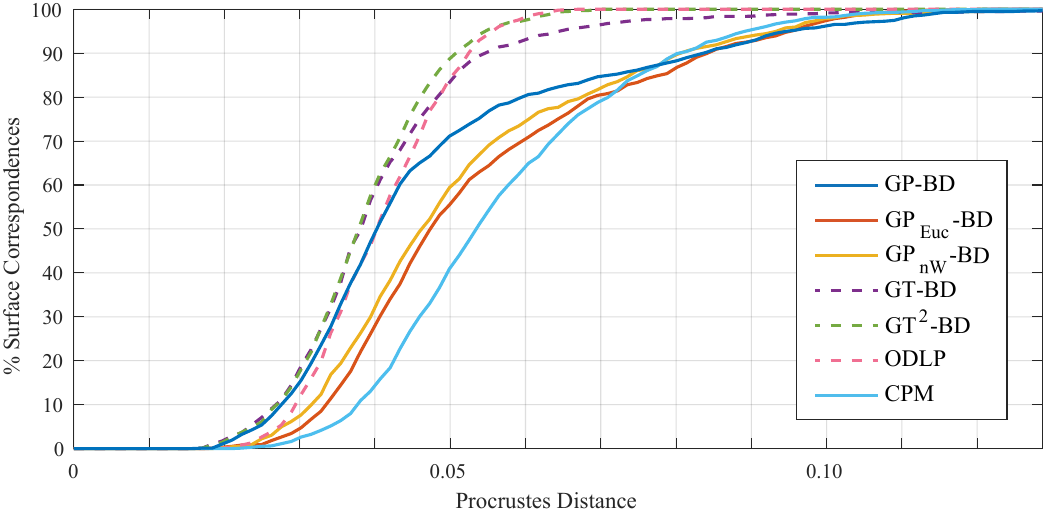}
  \caption{\small Comparing GP-BD to the alternative methods listed in \cref{table:experiment_abbv} for the dataset of distal radii of apes and humans. The general trend is consistent with our observation in \cref{fig:error_plot}: GT-BD outperforms CPM and is comparable to variants of Gaussian process landmarks with alternative kernel constructions, whiling falling short of, but not far from, GT$^2$-DB and ODLP which rely additionally on the oracle ``ground truth'' correspondences between observer landmarks.}
  \label{fig:error_plot_radius}
\end{figure}

\begin{figure}[htpb]
  \centering
  \includegraphics[width=.85\textwidth]{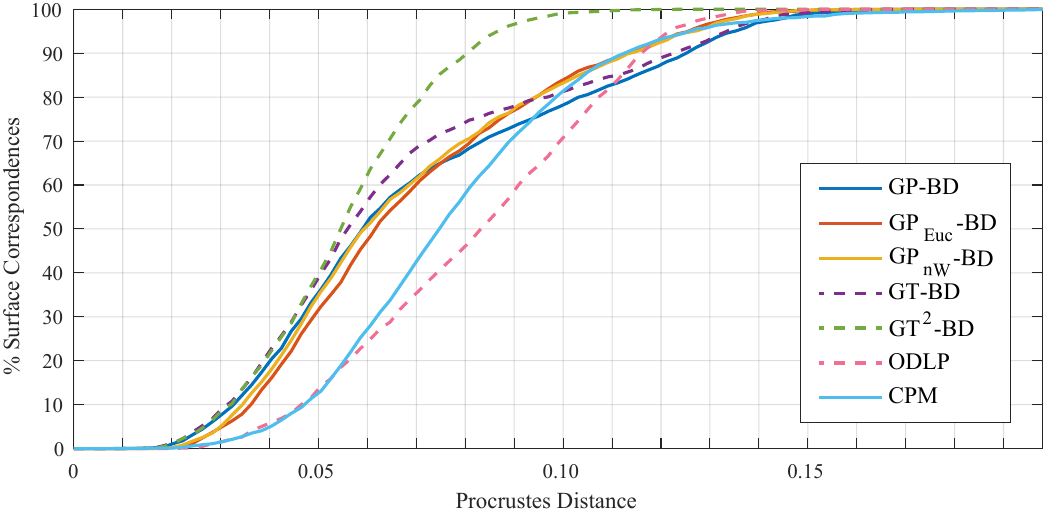}
  \caption{\small Comparing GP-BD to the alternative methods listed in \cref{table:experiment_abbv} for the dataset of first metatarsals of prosimian primates, New and Old World monkeys. The general trend is consistent with our observation in \cref{fig:error_plot}: GT-BD outperforms CPM and variants of Gaussian process landmarks with alternative kernel constructions, falls short of, but not far from, GT$^2$-DB and ODLP which rely additionally on the oracle ``ground truth'' correspondences between observer landmarks.}
  \label{fig:error_plot_mt1}
\end{figure}

In this section we present the results of the surface registration algorithm in \Cref{sec:surf-registr-match} on two other anatomical datasets, one consisting of $57$ proximal first metatarsals of prosimian primates, New and Old World monkeys (\cref{fig:error_plot_mt1}), and the other consisting of $45$ distal radii of apes and humans (\cref{fig:error_plot_radius}). Detailed descriptions of these two additional datasets can be found in \cite{PNAS2011}.

\begin{figure}[htb]
  \centering
  \includegraphics[width=1\textwidth]{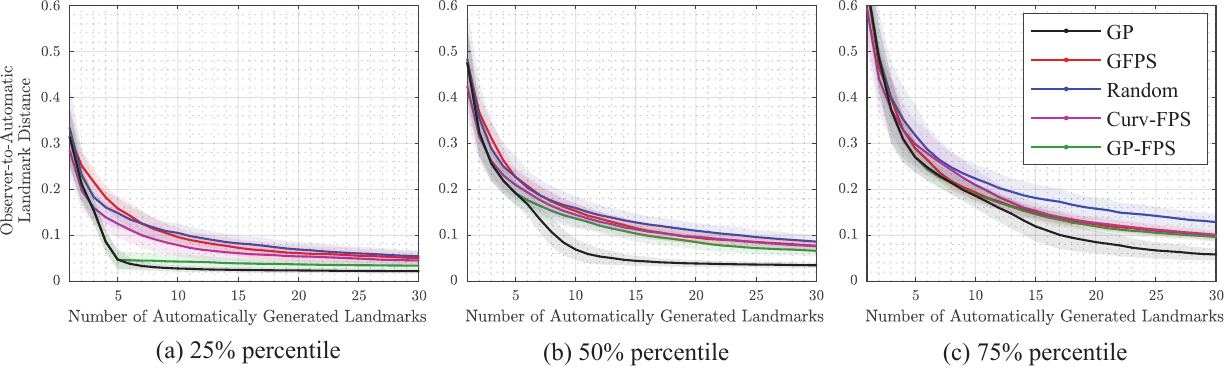}
  \caption{\small Geodesic distance from an observer landmark to its nearest automatically generated landmark, with respect to different numbers of automatic landmarks, on a collection of $116$ second mandibular molars of prosimian primates and closely related non-primates (see \cite{PNAS2011} for a more detailed description of this dataset and the observer landmark acquisition). Each point on any of the three solid curves is obtained by averaging the $25\%$, $50\%$ (median) and $75\%$ percentiles of observer-to-automatic landmark distances computed for each surface in this dataset; the transparent bands represent confidence intervals of one standard deviation. The types of automatic landmarks are Gaussian process landmarks (GP), geodesic farthest point sampling landmarks (GFPS), random landmarks uniformly selected from the vertices of each triangular mesh, and geodesic farthest point sampling landmarks initialized by either critical points of curvature (Curv-FPS) or the first few Gaussian process landmarks (GP-FPS). The random landmarks are only used as a baseline in this experiment.}
  \label{fig:compareCoverage-prctiles-teeth}
\end{figure}

\begin{figure}[htb]
  \centering
  \includegraphics[width=1.0\textwidth]{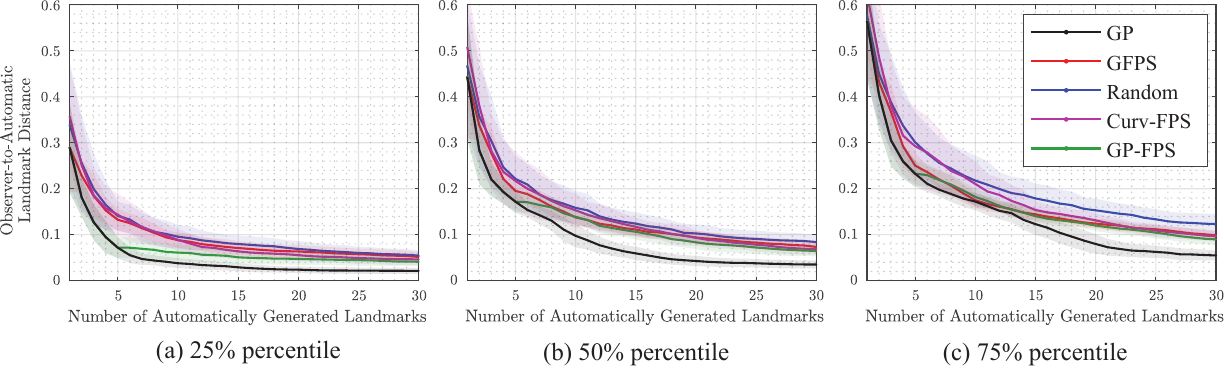}
  \caption{\small Geodesic distance from an observer landmark to its nearest automatically generated landmark, with respect to different numbers of automatic landmarks, on a collection of $45$ distal radii of apes and humans (see \cite{PNAS2011} for a more detailed description of this dataset and the observer landmark acquisition). Each point on any of the three solid curves is obtained by averaging the $25\%$, $50\%$ (median) and $75\%$ percentiles of observer-to-automatic landmark distances computed for each surface in this dataset; the transparent bands represent confidence intervals of one standard deviation. The types of automatic landmarks are Gaussian process landmarks (GP), geodesic farthest point sampling landmarks (GFPS), random landmarks uniformly selected from the vertices of each triangular mesh, and geodesic farthest point sampling landmarks initialized by either critical points of curvature (Curv-FPS) or the first few Gaussian process landmarks (GP-FPS). The random landmarks are only used as a baseline in this experiment.}
  \label{fig:compareCoverage-prctiles-radius}
\end{figure}

\begin{figure}[htb]
  \centering
  \includegraphics[width=1.0\textwidth]{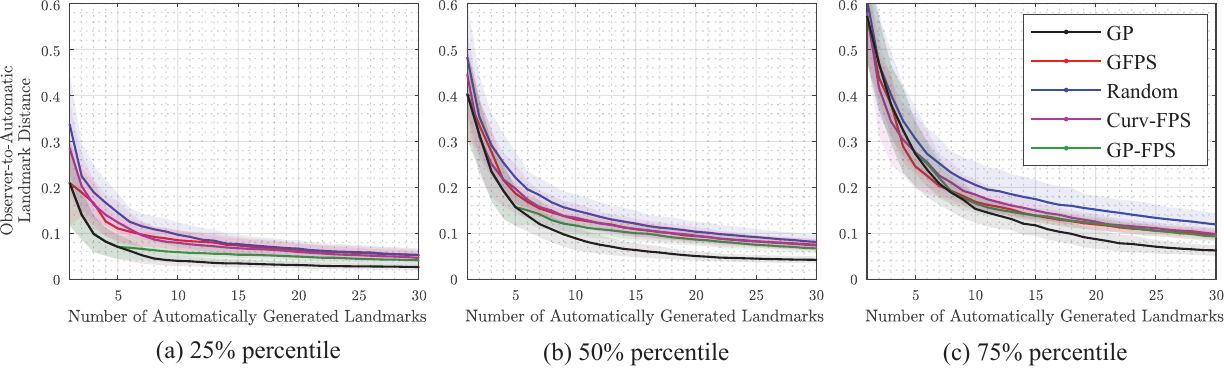}
  \caption{\small Geodesic distance from an observer landmark to its nearest automatically generated landmark, with respect to different numbers of automatic landmarks, on a collection of $57$ first metatarsals of prosimian primates (see \cite{PNAS2011} for a more detailed description of this dataset and the observer landmark acquisition). Each point on any of the three solid curves is obtained by averaging the $25\%$, $50\%$ (median) and $75\%$ percentiles of observer-to-automatic landmark distances computed for each surface in this dataset; the transparent bands represent confidence intervals of one standard deviation. The types of automatic landmarks are Gaussian process landmarks (GP), geodesic farthest point sampling landmarks (GFPS), random landmarks uniformly selected from the vertices of each triangular mesh, and geodesic farthest point sampling landmarks initialized by either critical points of curvature (Curv-FPS) or the first few Gaussian process landmarks (GP-FPS). The random landmarks are only used as a baseline in this experiment.}
  \label{fig:compareCoverage-prctiles-mt1}
\end{figure}

We provide additional figures (\cref{fig:compareCoverage-prctiles-teeth}, \cref{fig:compareCoverage-prctiles-radius}, \cref{fig:compareCoverage-prctiles-mt1}) illustrating the advantageous coverage properties of the Gaussian process landmarks, in terms of their geodesic distances to observer landmarks. Each figure corresponds to one of the three datasets (molars, first metatarsals, and radii) and contains three subplots --- one for each of three levels of quantile ($25\%$, $50\%$, $75\%$). See captions of these figures (also that of \cref{fig:compareCoverage-median}) for more detailed explanations.

\begin{figure}[htb]
  \centering
  \includegraphics[width=1.0\textwidth]{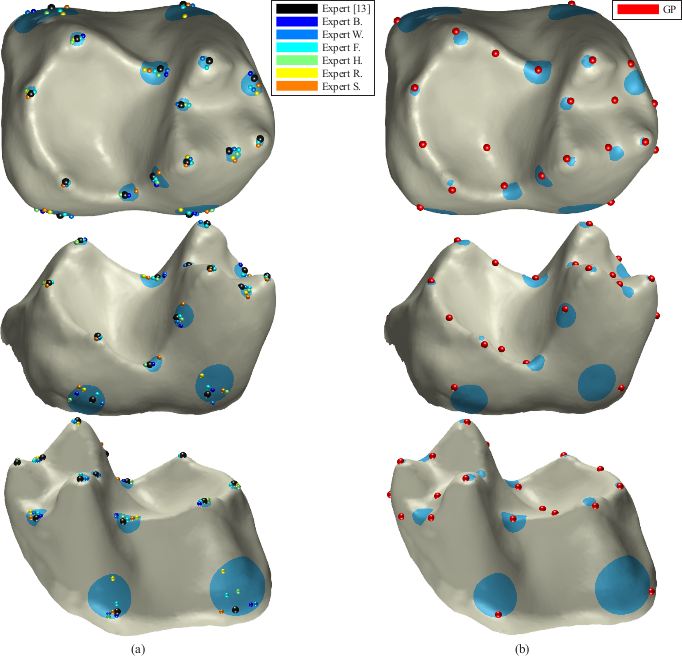}
  \caption{\small (a) Observer landmarks placed by multiple experts on a fossil molar. Landmarks obtained in \cite{PNAS2011}, shown in black color, were identified manually by an evolutionary anthropologist and used throughout this paper as ground truth. 6 additional sets of landmarks are shown in multiple colors; these landmarks were provided by experts guided to manually place landmarks following the same landmarking scheme, so as to capture the same morphological features; see Figure \cref{fig:Comp_UserLmksSupp_Experts} for a visualization of the landmarks provided by each individual expert. Cyan-shaded areas depict the smallest geodesic disk enclosing each group of observer landmarks. (b) Gaussian process landmarks computed with \cref{alg:gaussian-process-landmarking} are shown in red along the cyan disks representing the distribution of expert-place landmarks. This further exemplifies the potential of the first few Gaussian process landmarks in identifying biologically-informative feature points. This rows of this figure provide additional views of the example provided in \cref{fig:Comp_UserLmks}.}
  \label{fig:Comp_UserLmks_Supp}
\end{figure}

\begin{figure}[htb]
  \centering
  \includegraphics[width=1.0\textwidth]{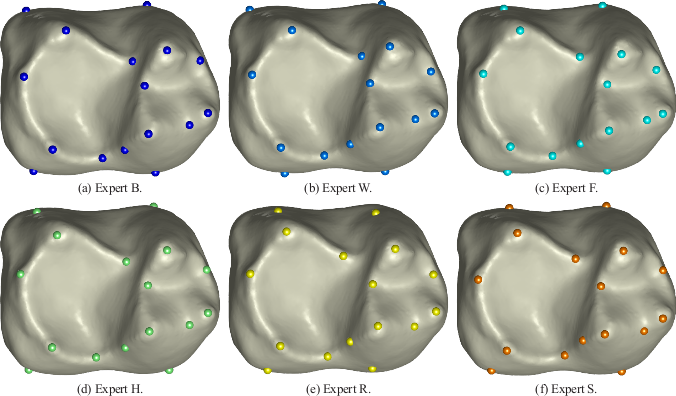}
  \caption{\small Multiple sets of observer landmarks placed by experts on a fossil molar. Independent experts were instructed to place landmarks following the protocol of \cite{PNAS2011} (pre-fixed ordering and/or number of landmarks of each type in the typology), the same protocol used for obtaining the landmarks used throughout this paper as ground truth. See \cref{fig:Comp_UserLmks} and \cref{fig:Comp_UserLmks_Supp} for a comparison of observer landmarks and Gaussian process landmarks automatically computed with \cref{alg:gaussian-process-landmarking}.}
  \label{fig:Comp_UserLmksSupp_Experts}
\end{figure}

We also provide in \cref{fig:Comp_UserLmks_Supp} and \cref{fig:Comp_UserLmksSupp_Experts} additional illustrations of multiple sets of expert landmarks manually placed following the same protocol in \cite{PNAS2011}. This is supplementary to \cref{fig:Comp_UserLmks} in the main text. We refer to the captions of the these figures for more detailed information.

\section{Ordination and Comparative Biological Interpretation for Shape Collections}
\label{sec:comp-bio-interp}

This section provides a qualitative but visually more straightforward way to compare the shape distance matrices obtained from the various pairwise registration methods in \cref{table:experiment_abbv} on a collection of primate molars. The results presented here suggests that shape distances obtained by matching Gaussian process landmarks can potentially capture more morphometrical information and provide better characterizations of the shape variation across species groups.

Comparative biologists, ecologists, and other natural and social scientists frequently rely on \emph{ordination} or \emph{gradient analysis} to summarize and emphasize patterns in multivariate datasets \cite{ClarkeGreen1988,Clarke1993,Anderson2001}. The equivalent but more familiar term for statisticians and applied mathematicians is \emph{dimension reduction}. Generally speaking, it is assumed or hypothesized that groups that are distantly related or have different ecologies will display some physical differences related to these variables. Nonetheless, it is difficult to know which physical traits reflect phylogenetic and ecological differences \emph{a priori}, or whether certain traits are independent of each other. Ordination methods such as \emph{Principal Components Analysis} (PCA), \emph{Multi-Dimensional Scaling} (MDS), and \emph{Correspondence Analysis} (CA) reveal the nature of those physical differences by summarizing multivariate datasets in lower dimensional graphical form.

\begin{figure}[htp]
\begin{subfigure}{.5\textwidth}
  \centering
  \includegraphics[width=1.0\linewidth]{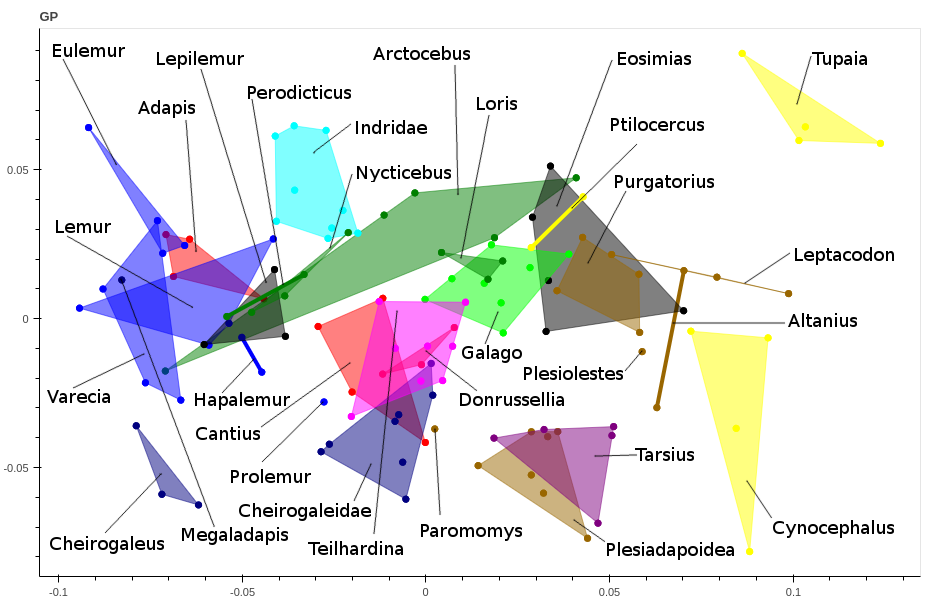}
  \caption{GP-BD}
  \label{fig:gp}
\end{subfigure}%
\begin{subfigure}{.5\textwidth}
  \centering
  \includegraphics[width=1.0\linewidth]{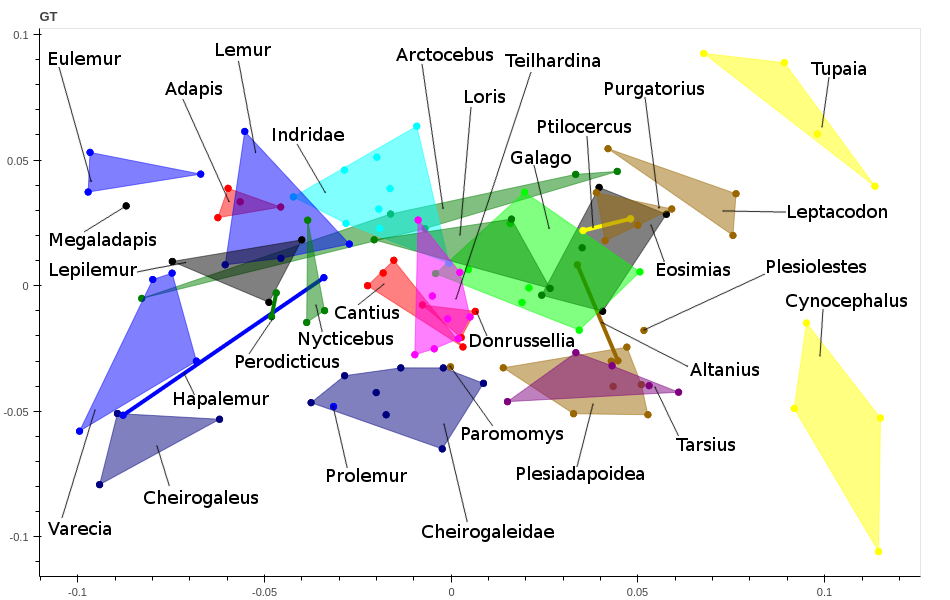}
  \caption{GT-BD}
  \label{fig:gt}
\end{subfigure}
\begin{subfigure}{.5\textwidth}
  \centering
  \includegraphics[width=1.0\linewidth]{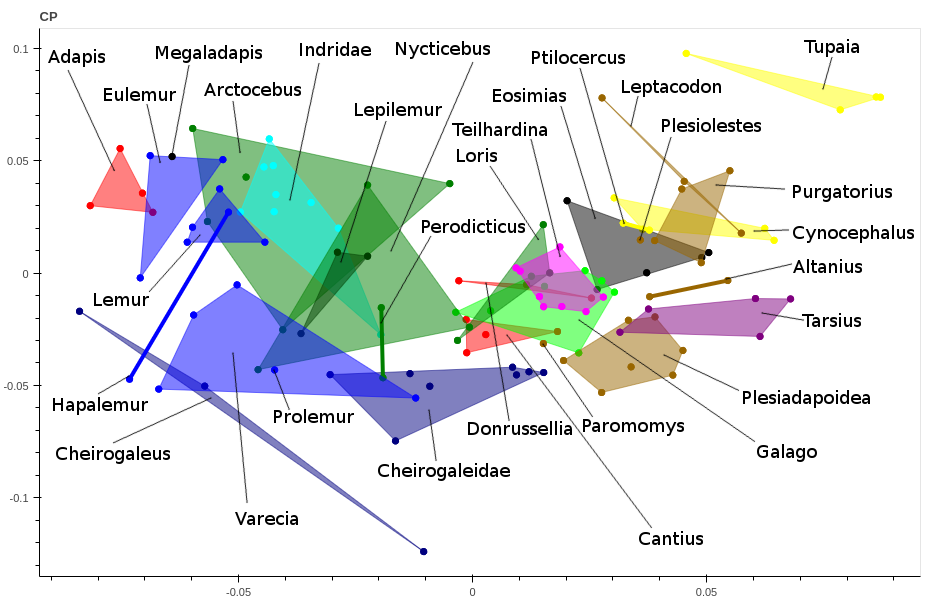}
  \caption{CPM}
  \label{fig:cp}
\end{subfigure}%
\begin{subfigure}{.5\textwidth}
  \centering
  \includegraphics[width=1.0\linewidth]{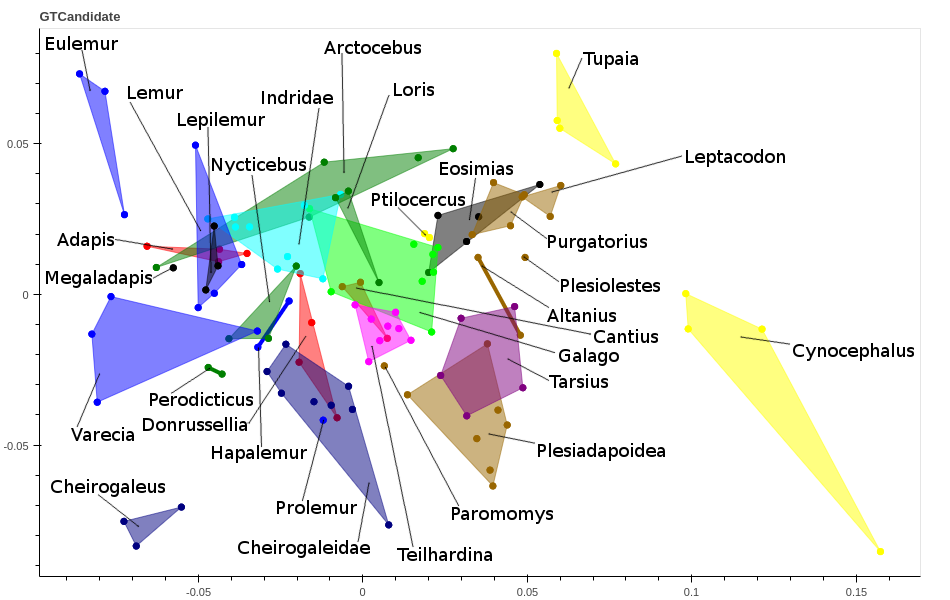}
  \caption{GT$^2$-BD}
  \label{fig:gtcandidate}
\end{subfigure}
\caption{\small Multi-Dimensional Scaling (MDS) plots visualizing shape similarities characterized by four alternative distance matrices for a collection of $116$ second mandibular molars of prosimian primates and non-primate close relatives. The distances investigated here include \textbf{(a)} Procrustes distance induced by GP-BD maps that match as many pairs of Gaussian process landmarks as possible within distortion bound $K=1.5$; \textbf{(b)} Procrustes distance induced by GT-BD maps that match as many pairs of observer placed landmarks as possible within distortion bound $K=1.5$, but without ``ground truth'' correspondences between observer placed landmarks; \textbf{(c)} continuous Procrustes distance \cite{CP13}; \textbf{(d)} Procrustes distance induced by GT$^2$-BD maps that match as many observed placed landmarks as possible within distortion bound $K=1.5$, with ``ground truth'' correspondences between observer placed landmarks as initial matching candidates. Each polygon encloses a collection of individual specimens of closely related species expected to be similar based on visual inspection and traditional comparative analyses. Similar ordination plots for GP$_{\textrm{Euc}}$ or GP$_{\textrm{nW}}$ are omitted in this figure because the ordination shown in those two plots are much worse than any of the plots (a)--(d). According to the PERMANOVA test results in \cref{table:permanova}, pairwise surface registration with GP-BD achieves comparable power of separating taxonomic groups as matching ``ground truth'' landmarks placed by professional comparative biologists.}
\label{fig:species-ordination-plots}
\end{figure}

In \cref{fig:species-ordination-plots} we visualize ordinations induced from four different distance matrices using classical multi-dimensional scaling (MDS). The dataset consists of $116$ second mandibular molars of prosimian primates and non-primate close relatives, first published in \cite{PNAS2011}. The distances include the continuous Procrustes distance \cite{CP13} and three BD-filtering methods; see \cref{table:experiment_abbv} for the abbreviations. In each ordination MDS plot, points corresponding to the same species group are enclosed in a polygonal, and the color pattern of the polygons indicates shape similarity among generic specimens across species groups, based on visual inspection and traditional comparative analyses. Similar ordination plots have been used for comparing different algorithms designed for automatically quantifying the geometric similarity of anatomical surfaces; see e.g., \cite{Auto3dGM2015,TMB2014,GYDMB2017}. This type of plots can also be considered as visualizations of \emph{morphospaces} in evolutionary and developmental biology \cite{MitteroeckerHuttegger2009}.

An important criterion for good ordination is the extent to which specimens belonging to the same species group cluster near each other, and specimens from different species groups are separated from each other. A popular summary statistic is to measure the ratio between within-group distances and among-group distances, which we quantitatively calculate and present in \cref{table:permanova}. Nevertheless, it is already evident from qualitatively inspecting \cref{fig:species-ordination-plots} that bounded distortion matching (\cref{fig:gp}, \cref{fig:gt}, \cref{fig:gtcandidate}) in general better distinguishes species groups than continuous Procrustes maps (\cref{fig:cp}). Considering taxonomic group membership and previous interpretations regarding shape affinities of different taxa, the ordination induced from GP-BD distances (\cref{fig:gp}) is visually most appealing. We explain some compelling examples in detail, referencing previous observations by biologists and paleontologists.
\begin{enumerate}[(i)]

\item\label{item:1} \cref{fig:gp} has a relatively low within/between-group distance ratio (see \cref{table:permanova} for more details), with remarkable distinctiveness of the Indriidae, \emph{Tupaia}, and \emph{Cynocephalus} groups from the other groups, which is a faithful reflection of the shape dissimilarity based on visual inspections;

\item\label{item:2} Among all different ordinations presented in \cref{fig:species-ordination-plots}, \cref{fig:gp} best separates out fossil (the brown-colored groups --- Plesiadapoidea, \emph{Plesiolestes}, \emph{Purgatorius}, \emph{Leptacodon}, and \emph{Altanius}) and living non-primates (the yellow-colored groups --- \emph{Tupaia} and \emph{Ptilocercus} (Scandentia or ``treeshrews'') and \emph{Cynocephalus} (Dermoptera)) from strepsirrhine primates and most early fossil primates (light and dark groups of lemurs --- \emph{Eulemur}, \emph{Lemur}, \emph{Varecai}, \emph{Cheirogaleus}, \emph{Hapalemur}, \emph{Prolemur}, and Cheirogaleidae; light and dark green groups of Lorisidae --- \emph{Loris}, \emph{Galago}, \emph{Arctocebus}, \emph{Nycticebus}, and \emph{Perodicticus}; the red group of the extinct adapiforms and omomyiforms --- \emph{Adapis}, \emph{Cantius}, and \emph{Donrussellia});

\item\label{item:5} The thought-provoking patterns of overlapping reflected in \cref{fig:gp} are in accordance with views established in existing comparative biological literature. For instance, \emph{Tarsius} (a haplorhine primate) and plesiadapoid nonprimates overlap, reflecting the idea by previous authors that they might be united in a common group called ``Plesiotarsiiformes'' \cite{Gidley1923,Gingerich1976}. As another example, \emph{Teilhardina}, \emph{Donrussellia}, and \emph{Cantius} all overlap, reflecting the fact that all of them represent very primitive members of primates: \emph{Donrussellia} was originally thought to be a new species of \emph{Teilhardina} \cite{RLS1967}; \emph{Cantius} was originally thought to be an omomyiform like \emph{Teilhardina} instead of an adapiform \cite{Simons1962}, and \cite{Gingerich1986} observed additional features of the premolar teeth uniquely relating \emph{Cantius torresi} to \emph{Teilhardina}. Though \emph{Cantius} is now reocognised as an adapiform, the wide separation between it and the other sampled adapiform (\emph{Adapis}) also matches previous qualitative discussions and analyses of independent datasets. In particular, the fact that \emph{Adapis} overlaps \emph{Lemur}, \emph{Eulemur}, and \emph{Lepilemur} is reminiscent of arguments by \cite{Gingerich1977} that \emph{Adapis} has special affinities to the living strepsirrhines to the exclusion of other adapiforms, as well as that \emph{Lepilemur} belongs to a group called ``Megaladapidae'' in part because of the claim that some members of the group have teeth extremely similar to those of \emph{Adapis} (\cite{SchwartzTattersall1985}). Anaylsis of other skeletal regions such as the ankle bones \cite{Seiffert2015} also link \emph{Adapis} more closely to living lemurs than to other adapiforms. Finally, the overlap of Eosimias with treeshrews matches the suggestion by \cite{Godinot2007} that this taxon, otherwise thought of as an anthropoid, is in fact treeshrew-like in its dentition.  Again analyses of ankle bones return a similar pattern; see e.g., \cite{Yapuncich2017,Boyer2017}.
\end{enumerate}

\bibliographystyle{siamplain}
\bibliography{references}

\begin{thebibliography}{100}

\bibitem{ARS2013}
{\sc D.~C. Adams, F.~J. Rohlf, and D.~E. Slice}, {\em {A Field Comes of Age:
  Geometric Morphometrics in the 21st Century}}, Hystrix, 24 (2013), pp.~7--14.

\bibitem{CP13}
{\sc R.~Al-Aifari, I.~Daubechies, and Y.~Lipman}, {\em {Continuous Procrustes
  Distance Between Two Surfaces}}, {C}ommunications on {P}ure and {A}pplied
  {M}athematics, 66 (2013), pp.~934--964,
  \url{https://doi.org/10.1002/cpa.21444},
  \url{http://dx.doi.org/10.1002/cpa.21444}.

\bibitem{ACDLD2003}
{\sc P.~Alliez, D.~Cohen-Steiner, O.~Devillers, B.~L\'{e}vy, and M.~Desbrun},
  {\em {Anisotropic Polygonal Remeshing}}, ACM Trans. Graph., 22 (2003),
  pp.~485--493, \url{https://doi.org/10.1145/882262.882296},
  \url{http://doi.acm.org/10.1145/882262.882296}.

\bibitem{Anderson2001}
{\sc M.~J. Anderson}, {\em {A New Method for Non-Parametric Multivariate
  Analysis of Variance}}, Austral Ecology, 26 (2001), pp.~32--46.

\bibitem{ASC2011}
{\sc M.~Aubry, U.~Schlickewei, and D.~Cremers}, {\em {The Wave Kernel
  Signature: A Quantum Mechanical Approach to Shape Analysis}}, in {Computer
  Vision Workshops (ICCV Workshops), 2011 IEEE International Conference on},
  IEEE, 2011, pp.~1626--1633.

\bibitem{Banchoff1970}
{\sc T.~F. Banchoff}, {\em {Critical Points and Curvature for Embedded
  Polyhedral Surfaces}}, The American Mathematical Monthly, 77 (1970),
  pp.~475--485.

\bibitem{Teilhardina2017}
{\sc C.~Beard}, {\em {Teilhardina. The International Encyclopedia of
  Primatology. 1--2.}}, 2017.
\newblock DOI: 10.1002/9781119179313.wbprim0444.

\bibitem{LapEigMaps2003}
{\sc M.~Belkin and P.~Niyogi}, {\em {Laplacian Eigenmaps for Dimensionality
  Reduction and Data Representation}}, {N}eural {C}omput., 15 (2003),
  pp.~1373--1396, \url{https://doi.org/10.1162/089976603321780317},
  \url{http://dx.doi.org/10.1162/089976603321780317}.

\bibitem{BerryHarlim2016}
{\sc T.~Berry and J.~Harlim}, {\em {Variable Bandwidth Diffusion Kernels}},
  Applied and Computational Harmonic Analysis, 40 (2016), pp.~68--96,
  \url{https://doi.org/10.1016/j.acha.2015.01.001},
  \url{http://www.sciencedirect.com/science/article/pii/S1063520315000020}.

\bibitem{BCDDvPW2011}
{\sc P.~Binev, A.~Cohen, W.~Dahmen, R.~DeVore, G.~Petrova, and P.~Wojtaszczyk},
  {\em {Convergence Rates for Greedy Algorithms in Reduced Basis Methods}},
  SIAM Journal on Mathematical Analysis, 43 (2011), pp.~1457--1472.

\bibitem{Bookstein1991}
{\sc F.~L. Bookstein}, {\em {Morphometric Tools for Landmark Data: Geometry and
  Biology}}, Cambridge University Press, 1991.

\bibitem{Boyer2012}
{\sc D.~M. Boyer, L.~Costeur, and Y.~Lipman}, {\em {Earliest Record of
  Platychoerops (Primates, Plesiadapidae), a new species from Mouras Quarry,
  Mont de Berru, France}}, {A}merican {J}ournal of {P}hysical {A}nthropology,
  149 (2012), pp.~329--346, \url{https://doi.org/10.1002/ajpa.22119},
  \url{http://dx.doi.org/10.1002/ajpa.22119}.

\bibitem{PNAS2011}
{\sc D.~M. Boyer, Y.~Lipman, E.~{St. Clair}, J.~Puente, B.~A. Patel,
  T.~Funkhouser, J.~Jernvall, and I.~Daubechies}, {\em {Algorithms to
  Automatically Quantify the Geometric Similarity of Anatomical Surfaces}},
  {P}roceedings of the {N}ational {A}cademy of {S}ciences, 108 (2011),
  pp.~18221--18226, \url{https://doi.org/10.1073/pnas.1112822108},
  \url{http://www.pnas.org/content/108/45/18221.abstract},
  \url{https://arxiv.org/abs/http://www.pnas.org/content/108/45/18221.full.pdf+html}.

\bibitem{Auto3dGM2015}
{\sc D.~M. Boyer, J.~Puente, J.~T. Gladman, C.~Glynn, S.~Mukherjee, G.~S.
  Yapuncich, and I.~Daubechies}, {\em {A New Fully Automated Approach for
  Aligning and Comparing Shapes}}, The Anatomical Record, 298 (2015),
  pp.~249--276.

\bibitem{Boyer2017}
{\sc D.~M. Boyer, S.~Toussaint, and M.~Godinot}, {\em {Postcrania of the Most
  Primitive Euprimate and Implications for Primate Origins}}, Journal of Human
  Evolution, 111 (2017), pp.~202--215.

\bibitem{Bronstein2011}
{\sc A.~M. Bronstein}, {\em {Spectral Descriptors for Deformable Shapes}},
  arXiv preprint arXiv:1110.5015,  (2011).

\bibitem{CCFM2008}
{\sc U.~Castellani, M.~Cristani, S.~Fantoni, and V.~Murino}, {\em {Sparse
  Points Matching by Combining 3D Mesh Saliency with Statistical Descriptors}},
  Computer Graphics Forum, 27 (2008), pp.~643--652.

\bibitem{CKS2015}
{\sc K.~N. Chaudhury, Y.~Khoo, and A.~Singer}, {\em {{G}lobal {R}egistration of
  {M}ultiple {P}oint {C}louds using {S}emidefinite {P}rogramming}}, {SIAM}
  {J}ournal on {O}ptimization, 25 (2015), pp.~468--501.

\bibitem{ClarkeGreen1988}
{\sc K.~Clarke and R.~Green}, {\em {Statistical Design and Analysis for a
  'Biological Effects' Study}}, Marine Ecology Progress Series,  (1988),
  pp.~213--226.

\bibitem{Clarke1993}
{\sc K.~R. Clarke}, {\em {Non-Parametric Multivariate Analyses of Changes in
  Community Structure}}, Austral Ecology, 18 (1993), pp.~117--143.

\bibitem{CM2003}
{\sc D.~Cohen-Steiner and J.-M. Morvan}, {\em {Restricted Delaunay
  Triangulations and Normal Cycle}}, in {Proceedings of the nineteenth annual
  symposium on Computational geometry}, ACM, 2003, pp.~312--321.

\bibitem{CoifmanLafon2006}
{\sc R.~R. Coifman and S.~Lafon}, {\em {Diffusion Maps}}, {A}pplied and
  {C}omputational {H}armonic {A}nalysis, 21 (2006), pp.~5--30,
  \url{https://doi.org/10.1016/j.acha.2006.04.006},
  \url{http://www.sciencedirect.com/science/article/pii/S1063520306000546}.
\newblock Special Issue: Diffusion Maps and Wavelets.

\bibitem{CzajaEhler2013}
{\sc W.~Czaja and M.~Ehler}, {\em {Schroedinger Eigenmaps for the Analysis of
  Biomedical Data}}, IEEE Transactions on Pattern Analysis and Machine
  Intelligence, 35 (2013), pp.~1274--1280.

\bibitem{DvPW2013}
{\sc R.~DeVore, G.~Petrova, and P.~Wojtaszczyk}, {\em {Greedy Algorithms for
  Reduced Bases in Banach Spaces}}, Constructive Approximation, 37 (2013),
  pp.~455--466.

\bibitem{DrydenMardia1998SSA}
{\sc I.~L. Dryden and K.~V. Mardia}, {\em {Statistical Shape Analysis}},
  vol.~4, John Wiley \& Sons New York, 1998.

\bibitem{FryThesis1993}
{\sc D.~S. Fry}, {\em {{S}hape {R}ecognition {U}sing {M}etrics on the {S}pace
  of {S}hapes}}, PhD thesis, Harvard University, Cambridge, MA, USA, 1993.
\newblock UMI Order No. GAX94-12337.

\bibitem{Gao2015Thesis}
{\sc T.~Gao}, {\em {Hypoelliptic Diffusion Maps and Their Applications in
  Automated Geometric Morphometrics}}, PhD thesis, Duke University, 2015.

\bibitem{HDM2016}
{\sc T.~Gao}, {\em {The Diffusion Geometry of Fibre Bundles}}, submitted,
  (2016).
\newblock arXiv:1602.02330.

\bibitem{GBM2016}
{\sc T.~Gao, J.~Brodzki, and S.~Mukherjee}, {\em {The Geometry of
  Synchronization Problems and Learning Group Actions}}, arXiv preprint
  arXiv:1610.09051,  (2016).

\bibitem{GPLMK1}
{\sc T.~Gao, S.~Z. Kovalsky, and I.~Daubechies}, {\em {Gaussian Process
  Landmarking on Manifolds}}, SIAM Journal on Mathematics of Data Science,
  (2019), \url{https://arxiv.org/abs/1802.03479}.
\newblock to appear.

\bibitem{GYDMB2017}
{\sc T.~Gao, G.~S. Yapuncich, I.~Daubechies, S.~Mukherjee, and D.~M. Boyer},
  {\em {Development and Assessment of Fully Automated and Globally Transitive
  Geometric Morphometric Methods, With Application to a Biological Comparative
  Dataset With High Interspecific Variation}}, The Anatomical Record: Advances
  in Integrative Anatomy and Evolutionary Biology,  (2017), pp.~636--658.

\bibitem{PCM2014}
{\sc L.~Z. Garamszegi}, {\em {Modern Phylogenetic Comparative Methods and Their
  Application in Evolutionary Biology}}, Concepts and Practice. London, UK:
  Springer,  (2014).

\bibitem{Gidley1923}
{\sc J.~W. Gidley}, {\em {Paleocene Primates of the Fort Union, with Discussion
  of Relationships of Eocene Primates}}, Proceedings of the United States
  National Museum, 63 (1923), pp.~1--38.

\bibitem{Gingerich1976}
{\sc P.~D. Gingerich}, {\em {Cranial Anatomy and Evolution of Early Tertiary
  Plesiadapidae (Mammalia, Primates)}}, University of Michigan Papers on
  Paleontology, 15 (1976), pp.~1--141.

\bibitem{Gingerich1977}
{\sc P.~D. Gingerich}, {\em {Radiation of Eocene adapidae in Europe}}, Geobios,
  10 (1977), pp.~165--182.

\bibitem{Gingerich1986}
{\sc P.~D. Gingerich}, {\em {Early Eocene Cantius torresi---oldest Primate of
  Modern Aspect from North America}}, Nature, 319 (1986), pp.~319--321.

\bibitem{Godinot2007}
{\sc M.~Godinot}, {\em {Primate Origins: A Reappraisal of Historical Data
  Favoring Tupaiid Affinities}}, Primate Origins: Adaptations and Evolution.
  Springer, New York,  (2007), pp.~83--142.

\bibitem{Gonzalez1985}
{\sc T.~F. Gonzalez}, {\em {Clustering to Minimize the Maximum Intercluster
  Distance}}, Theoretical Computer Science, 38 (1985), pp.~293--306.

\bibitem{Good2004}
{\sc P.~I. Good}, {\em {Permutation, Parametric, and Bootstrap Tests of
  Hypotheses (Springer Series in Statistics)}}, Springer-Verlag New York, Inc.,
  Secaucus, NJ, USA, 2004.

\bibitem{Gower1975}
{\sc J.~C. Gower}, {\em {Generalized Procrustes Analysis}}, Psychometrika, 40
  (1975), pp.~33--51, \url{https://doi.org/10.1007/BF02291478}.

\bibitem{GowerDijksterhuis2004GPA}
{\sc J.~C. Gower and G.~B. Dijksterhuis}, {\em {Procrustes Problems}}, vol.~3
  of {Oxford Statistical Science Series}, Oxford University Press Oxford, 2004.

\bibitem{Gueneysu2010}
{\sc B.~G\"{u}neysu}, {\em {The Feynman-Kac Formula for Schr\"{o}dinger
  Operators on Vector Bundles over Complete Manifolds}}, Journal of Geometry
  and Physics, 60 (2010), pp.~1997--2010.

\bibitem{HKVHMJ2012}
{\sc E.~Harjunmaa, A.~Kallonen, M.~Voutilainen, K.~H\"{a}m\"{a}l\"{a}inen,
  M.~L. Mikkola, and J.~Jernvall}, {\em {On the Difficulty of Increasing Dental
  Complexity}}, Nature, 483 (2012), pp.~324--327.

\bibitem{HSHRCKZEMS2014}
{\sc E.~Harjunmaa, K.~Seidel, T.~H\"{a}kkinen, E.~Renvois\'{e}, I.~J. Corfe,
  A.~Kallonen, Z.-Q. Zhang, A.~R. Evans, M.~L. Mikkola, I.~Salazar-Ciudad,
  et~al.}, {\em {Replaying Evolutionary Transitions From the Dental Fossil
  Record}}, Nature, 512 (2014), pp.~44--48.

\bibitem{HLB2016}
{\sc B.~R. Hassett and T.~Lewis-Bale}, {\em {Comparison of 3D Landmark and 3D
  Dense Cloud Approaches to Hominin Mandible Morphometrics Using
  Structure-From-Motion}}, Archaeometry, 59 (2017), pp.~191--203,
  \url{https://doi.org/10.1111/arcm.12229},
  \url{http://dx.doi.org/10.1111/arcm.12229}.
\newblock ARCH-05-0070-2015.R2.

\bibitem{HS1985}
{\sc B.~Helffer and J.~Sj\"{o}strand}, {\em {Puits Multiples en Mecanique
  Semi-Classique iv Etude du Complexe de Witten}}, Communications in partial
  differential equations, 10 (1985), pp.~245--340.

\bibitem{JoshiMiller2000}
{\sc S.~C. Joshi and M.~I. Miller}, {\em {Landmark Matching via Large
  Deformation Diffeomorphisms}}, {IEEE Transactions on Image Processing}, 9
  (2000), pp.~1357--1370, \url{https://doi.org/10.1109/83.855431}.

\bibitem{Kendall1984}
{\sc D.~G. Kendall}, {\em {{S}hape {M}anifolds, {P}rocrustean {M}etrics, and
  {C}omplex {P}rojective {S}paces}}, {B}ulletin of the {L}ondon {M}athematical
  {S}ociety, 16 (1984), pp.~81--121.

\bibitem{KLF2011}
{\sc V.~Kim, Y.~Lipman, and T.~Funkhouser}, {\em {Blended Intrinsic Maps}}, ACM
  Transactions on Graphics (Proc. SIGGRAPH), 30 (2011).

\bibitem{KLQ1995}
{\sc C.-W. Ko, J.~Lee, and M.~Queyranne}, {\em {An Exact Algorithm for Maximum
  Entropy Sampling}}, Operations Research, 43 (1995), pp.~684--691.

\bibitem{KoehlHass2015}
{\sc P.~Koehl and J.~Hass}, {\em {Landmark-Free Geometric Methods in Biological
  Shape Analysis}}, Journal of The Royal Society Interface, 12 (2015),
  p.~20150795.

\bibitem{kovalsky2016}
{\sc S.~Z. Kovalsky, M.~Galun, and Y.~Lipman}, {\em {Accelerated quadratic
  proxy for geometric optimization}}, ACM Transactions on Graphics (TOG), 35
  (2016), p.~134.

\bibitem{KSG2008}
{\sc A.~Krause, A.~Singh, and C.~Guestrin}, {\em {Near-Optimal Sensor
  Placements in Gaussian Processes: Theory, Efficient Algorithms and Empirical
  Studies}}, Journal of Machine Learning Research, 9 (2008), pp.~235--284.

\bibitem{Lipman2012}
{\sc Y.~Lipman}, {\em {Bounded Distortion Mapping Spaces for Triangular
  Meshes}}, ACM Transactions on Graphics (TOG), 31 (2012), p.~108.

\bibitem{LipmanDaubechies2011}
{\sc Y.~Lipman and I.~Daubechies}, {\em {Conformal Wasserstein Distances:
  Comparing Surfaces in Polynomial Time}}, {A}dvances in {M}athematics, 227
  (2011), pp.~1047--1077, \url{https://doi.org/10.1016/j.aim.2011.01.020},
  \url{http://www.sciencedirect.com/science/article/pii/S0001870811000351}.

\bibitem{LipmanPuenteDaubechies2013}
{\sc Y.~Lipman, J.~Puente, and I.~Daubechies}, {\em {Conformal Wasserstein
  Distance: II. Computational Aspects and Extensions.}}, {M}ath. {C}omput., 82
  (2013), \url{http://dblp.uni-trier.de/db/journals/moc/moc82.html#LipmanPD13}.

\bibitem{LYPJB2014}
{\sc Y.~Lipman, S.~Yagev, R.~Poranne, D.~W. Jacobs, and R.~Basri}, {\em
  {Feature Matching with Bounded Distortion}}, ACM Transactions on Graphics
  (TOG), 33 (2014), p.~26.

\bibitem{LLKR2007}
{\sc Y.-S. Liu, M.~Liu, D.~Kihara, and K.~Ramani}, {\em {Salient Critical
  Points for Meshes}}, in {Proceedings of the 2007 ACM symposium on Solid and
  physical modeling}, ACM, 2007, pp.~277--282.

\bibitem{Lowe1999}
{\sc D.~G. Lowe}, {\em {Object Recognition from Local Scale-Invariant
  Features}}, in Computer vision, 1999. The proceedings of the seventh IEEE
  international conference on, vol.~2, Ieee, 1999, pp.~1150--1157.

\bibitem{Mantel1967}
{\sc N.~Mantel}, {\em {The Detection of Disease Clustering and a Generalized
  Regression Approach}}, Cancer research, 27 (1967), pp.~209--220.

\bibitem{MRCB2017}
{\sc S.~Melzi, E.~Rodol\`{a}, U.~Castellani, and M.~M. Bronstein}, {\em
  {Localized Manifold Harmonics for Spectral Shape Analysis}}, in {Computer
  Graphics Forum}, Wiley Online Library, 2017.

\bibitem{MG2009}
{\sc P.~Mitteroecker and P.~Gunz}, {\em {Advances in Geometric Morphometrics}},
  Evolutionary Biology, 36 (2009), pp.~235--247.

\bibitem{MitteroeckerHuttegger2009}
{\sc P.~Mitteroecker and S.~M. Huttegger}, {\em {The Concept of Morphospaces in
  Evolutionary and Developmental Biology: Mathematics and Metaphors}},
  Biological Theory, 4 (2009), pp.~54--67.

\bibitem{MoenningDodgson2003}
{\sc C.~Moenning and N.~A. Dodgson}, {\em {Fast Marching Farthest Point
  Sampling}}, tech. report, University of Cambridge, Computer Laboratory, 2003.

\bibitem{LeMoigne2017}
{\sc J.~L. Moigne}, {\em Introduction to remote sensing image registration}, in
  2017 IEEE International Geoscience and Remote Sensing Symposium (IGARSS),
  July 2017, pp.~2565--2568, \url{https://doi.org/10.1109/IGARSS.2017.8127519}.

\bibitem{NLCK2006}
{\sc B.~Nadler, S.~Lafon, R.~R. Coifman, and I.~G. Kevrekidis}, {\em {Diffusion
  Maps, Spectral Clustering and Reaction Coordinates of Dynamical Systems}},
  Applied and Computational Harmonic Analysis, 21 (2006), pp.~113--127.

\bibitem{NaorRegevVidick2013}
{\sc A.~Naor, O.~Regev, and T.~Vidick}, {\em {Efficient Rounding for the
  Noncommutative Grothendieck Inequality}}, in {Proceedings of the forty-fifth
  annual ACM symposium on Theory of computing}, ACM, 2013, pp.~71--80.

\bibitem{NWF1978}
{\sc G.~L. Nemhauser, L.~A. Wolsey, and M.~L. Fisher}, {\em {An Analysis of
  Approximations for Maximizing Submodular Set Functions-I}}, Mathematical
  Programming, 14 (1978), pp.~265--294.

\bibitem{Nemirovski2007}
{\sc A.~Nemirovski}, {\em {Sums of Random Symmetric Matrices and Quadratic
  Optimization under Orthogonality Constraints}}, {M}athematical programming,
  109 (2007), pp.~283--317.

\bibitem{NicholsHolmes2002}
{\sc T.~E. Nichols and A.~P. Holmes}, {\em {Nonparametric Permutation Tests for
  Functional Neuroimaging: A Primer with Examples}}, Human brain mapping, 15
  (2002), pp.~1--25.

\bibitem{SKKS2010}
{\sc S.~Niranjan, A.~Krause, S.~M. Kakade, and M.~Seeger}, {\em {Gaussian
  Process Optimization in the Bandit Setting: No Regret and Experimental
  Design}}, in {Proceedings of the 27th International Conference on Machine
  Learning}, 2010.

\bibitem{OHG2011}
{\sc M.~Ovsjanikov, Q.-X. Huang, and L.~Guibas}, {\em {A Condition Number for
  Non-Rigid Shape Matching}}, Computer Graphics Forum, 30 (2011),
  pp.~1503--1512.

\bibitem{OSS2018}
{\sc O.~{\"O}zye\c{s}il, N.~Sharon, and A.~Singer}, {\em {Synchronization Over
  Cartan Motion Groups via Contraction}}, SIAM Journal on Applied Algebra and
  Geometry, 2 (2018), pp.~207--241.

\bibitem{Pandey2009}
{\sc S.~Pandey, W.~Voorsluys, M.~Rahman, R.~Buyya, J.~Dobson, and K.~Chiu},
  {\em {Brain Image Registration Analysis Workflow for fMRI Studies on Global
  Grids}}, in 2009 International Conference on Advanced Information Networking
  and Applications, May 2009, pp.~435--442,
  \url{https://doi.org/10.1109/AINA.2009.13}.

\bibitem{Paradis2011}
{\sc E.~Paradis}, {\em {Analysis of Phylogenetics and Evolution with R}},
  Springer Science \& Business Media, 2011.

\bibitem{Pesarin2001}
{\sc F.~Pesarin}, {\em {Multivariate Permutation Tests: with Applications in
  Biostatistics}}, vol.~240, Wiley Chichester, 2001.

\bibitem{LPNV2013}
{\sc D.~L. Peutrec, F.~Nier, and C.~Viterbo}, {\em {Precise Arrhenius Law for
  p-Forms: The Witten Laplacian and Morse-Barannikov Complex}}, Annales Henri
  Poincar\'{e}, 14 (2013), pp.~567--610.

\bibitem{PEKGJ2008}
{\sc I.~Plyusnin, A.~R. Evans, A.~Karme, A.~Gionis, and J.~Jernvall}, {\em
  {Automated 3D Phenotype Analysis Using Data Mining}}, PLoS One, 3 (2008),
  p.~e1742.

\bibitem{PuenteThesis2013}
{\sc J.~Puente}, {\em {Distances and Algorithms to Compare Sets of Shapes for
  Automated Biological Morphometrics}}, PhD thesis, Princeton University, 2013.

\bibitem{RamsaySilverman2005}
{\sc J.~Ramsay and B.~Silverman}, {\em {Functional Data Analysis}}, {Springer
  Series in Statistics}, Springer, 2005.

\bibitem{RamsaySilverman2002}
{\sc J.~O. Ramsay and B.~W. Silverman}, {\em {Applied Functional Data Analysis:
  Methods and Case Studies}}, vol.~77, Springer New York, 2002.

\bibitem{GMBlueBook1990}
{\sc F.~J. Rohlf and F.~L. Bookstein}, {\em {Proceedings of the Michigan
  Morphometrics Workshop}}, University of Michigan Museum of Zoology, 1990.

\bibitem{Roth1993}
{\sc V.~L. Roth}, {\em {On Three-Dimensional Morphometrics, and on the
  Identification of Landmark Points}}, in {Contributions to Morphometrics},
  L.~F. Marcus, E.~Bello, and G.-V. A., eds., Museo Nacional de Ciencias
  Naturales, Madrid, 1993, pp.~41--61.

\bibitem{RLS1967}
{\sc D.~E. Russell, P.~Louis, and D.~E. Savage}, {\em {Primates of the French
  Early Eocene}}, University of California Publications in the Geological
  Sciences, 73 (1967), pp.~1--46.

\bibitem{SWN2013}
{\sc T.~J. Santner, B.~J. Williams, and W.~I. Notz}, {\em {The Design and
  Analysis of Computer Experiments}}, {Springer Series in Statistics}, Springer
  Science \& Business Media, 2013.

\bibitem{SchwartzTattersall1985}
{\sc J.~H. Schwartz and I.~Tattersall}, {\em {Evolutionary Relationships of
  Living Lemurs and Lorises (Mammalia, Primates) and Their Potential Affinities
  with European Eocene Adapidae.}}, Anthropological papers of the AMNH; v. 60,
  pt. 1,  (1985).

\bibitem{Seiffert2015}
{\sc E.~R. Seiffert, L.~Costeur, and D.~M. Boyer}, {\em {Tarsal Morphology of
  Caenopithecus, a Large Adapiform Primate from the Middle Eocene of
  Switzerland}}, PeerJ, 3 (2015), p.~e1036.

\bibitem{Simons1962}
{\sc E.~L. Simons}, {\em {A New Eocene Primate Genus, Cantius, and a Revision
  of Some Allied European Lemuroids}}, vol.~7, British Museum, 1962.

\bibitem{Singer2006ConvergenceRate}
{\sc A.~Singer}, {\em {From Graph to Manifold Laplacian: The Convergence
  Rate}}, {A}pplied and {C}omputational {H}armonic {A}nalysis, 21 (2006),
  pp.~128--134.

\bibitem{SingerWu2012VDM}
{\sc A.~Singer and H.-T. Wu}, {\em {Vector Diffusion Maps and the Connection
  Laplacian}}, {C}ommunications on {P}ure and {A}pplied {M}athematics, 65
  (2012), pp.~1067--1144, \url{https://doi.org/10.1002/cpa.21395},
  \url{http://dx.doi.org/10.1002/cpa.21395}.

\bibitem{smith2015}
{\sc J.~Smith and S.~Schaefer}, {\em {Bijective parameterization with free
  boundaries}}, ACM Transactions on Graphics (TOG), 34 (2015), p.~70.

\bibitem{SWW2000}
{\sc O.~Smolyanov, H.~Weizs\"{a}cker, and O.~Wittich}, {\em {Brownian Motion on
  a Manifold as Limit of Stepwise Conditioned Standard Brownian Motions}},
  Stochastic Processes, Physics and Geometry: New Interplays, II, 29 (2000),
  pp.~589--602.

\bibitem{SWW2007}
{\sc O.~G. Smolyanov, H.~v~Weizs\"{a}cker, and O.~Wittich}, {\em {Chernoff's
  Theorem and Discrete Time Approximations of Brownian Motion on Manifolds}},
  Potential Analysis, 26 (2007), pp.~1--29.

\bibitem{So2011}
{\sc A.~M.-C. So}, {\em {Moment Inequalities for Sums of Random Matrices and
  Their Applications in Optimization}}, {M}athematical {P}rogramming, 130
  (2011), pp.~125--151.

\bibitem{Stein2012}
{\sc M.~L. Stein}, {\em {Interpolation of Spatial Data: Some Theory for
  Kriging}}, Springer Science \& Business Media, 2012.

\bibitem{TCLLLMMSR2013}
{\sc G.~K. Tam, Z.-Q. Cheng, Y.-K. Lai, F.~C. Langbein, Y.~Liu, D.~Marshall,
  R.~R. Martin, X.-F. Sun, and P.~L. Rosin}, {\em {Registration of 3D Point
  Clouds and Meshes: A Survey from Rigid to Nonrigid}}, IEEE Transactions on
  Visualization and Computer Graphics, 19 (2013), pp.~1199--1217.

\bibitem{TMB2014}
{\sc K.~Turner, S.~Mukherjee, and D.~M. Boyer}, {\em {Persistent homology
  transform for modeling shapes and surfaces}}, Information and Inference: A
  Journal of the IMA, 3 (2014), pp.~310--344,
  \url{https://doi.org/10.1093/imaiai/iau011},
  \url{http://dx.doi.org/10.1093/imaiai/iau011}.

\bibitem{vanKaick2011}
{\sc O.~Van~Kaick, H.~Zhang, G.~Hamarneh, and D.~Cohen-Or}, {\em {A Survey on
  Shape Correspondence}}, Computer Graphics Forum, 30 (2011), pp.~1681--1707.

\bibitem{VMGBSB2017}
{\sc N.~Vitek, C.~Manz, T.~Gao, J.~Bloch, S.~Strait, and D.~M. Boyer}, {\em
  {Semi-supervised Determination of Pseudocryptic Morphotypes Using
  Observer-free Characterizations of Anatomical Alignment and Shape}}.
\newblock Accepted, 2017.

\bibitem{WGGPS2018}
{\sc S.~K. W{\"a}rml{\"a}nder, H.~Garvin, P.~Guyomarc'h, A.~Petaros, and S.~B.
  Sholts}, {\em {Landmark Typology in Applied Morphometrics Studies: What's the
  Point?}}, The Anatomical Record,  (2018).

\bibitem{Watanabe2018}
{\sc A.~Watanabe}, {\em {How Many Landmarks are Enough to Characterize Shape
  and Size Variation?}}, PloS one, 13 (2018), p.~e0198341.

\bibitem{White2009}
{\sc J.~White}, {\em {Geometric Morphometric Investigation of Molar Shape
  Diversity in Modern Lemurs and Lorises}}, The Anatomical Record: Advances in
  Integrative Anatomy and Evolutionary Biology: Advances in Integrative Anatomy
  and Evolutionary Biology, 292 (2009), pp.~701--719.

\bibitem{Witten1982}
{\sc E.~Witten}, {\em {Supersymmetry and Morse Theory}}, J. Differential Geom.,
  17 (1982), pp.~661--692, \url{https://doi.org/10.4310/jdg/1214437492},
  \url{https://doi.org/10.4310/jdg/1214437492}.

\bibitem{Yapuncich2017}
{\sc G.~S. Yapuncich, E.~R. Seiffert, and D.~M. Boyer}, {\em {Quantification of
  the Position and Depth of the Flexor Hallucis Longus Groove in Euarchontans,
  with Implications for the Evolution of Primate Positional Behavior}},
  American Journal of Physical Anthropology, 163 (2017), pp.~367--406.

\bibitem{ZSS2012}
{\sc M.~L. Zelditch, D.~L. Swiderski, and H.~D. Sheets}, {\em {Geometric
  Morphometrics for Biologists: A Primer}}, Academic Press, San Diego,
  second~ed., 2012.

\end{thebibliography}

\end{document}